\documentclass[preprint,12pt]{elsarticle}

\newif\ifshownotes
\shownotesfalse

\usepackage[utf8]{inputenc}
\usepackage{amsmath,amssymb}
\usepackage{amsthm}
\newtheorem{proposition}{Proposition}
\theoremstyle{definition}
\newtheorem{definition}{Definition}
\newtheorem{remark}{Remark}
\theoremstyle{plain}
\usepackage{graphicx}
\usepackage{placeins}
\usepackage{booktabs}
\usepackage{array}
\usepackage{multirow}
\usepackage{tabularx}
\usepackage{algorithm}
\usepackage{algorithmic}
\usepackage{xcolor}
\usepackage{lineno}
\usepackage{microtype}
\usepackage[hyphens]{url}
\usepackage[colorlinks=true,linkcolor=blue,citecolor=blue,urlcolor=blue]{hyperref}

\usepackage{enumitem}
\setlist{nosep}
\setlist[enumerate]{leftmargin=*}
\setlist[itemize]{leftmargin=*}

\usepackage[nolist]{acronym}
\AtBeginDocument{%
  \pdfstringdefDisableCommands{%
    \def\ac#1{#1}%
    \def\acp#1{#1s}%
    \def\acs#1{#1}%
  }%
}

\ifshownotes
  \newcommand{\Rami}[1]{\textbf{\color{red}[[Rami: #1]]}}
  \newcommand{\Aviram}[1]{\textbf{\color{brown}[[Aviram: #1]]}}
  \newcommand{\MergeNote}[1]{\textbf{\color{red}[[Merge check: #1]]}}
  \newcommand{\Eyal}[1]{\textbf{\color{blue}[[Eyal: #1]]}}
  \newcommand{\Shaun}[1]{\textbf{\color{magenta}[[Shaun: #1]]}}
  \newcommand{\Chen}[1]{\textbf{\color{teal}[[Chen: #1]]}}
  \newcommand{\inderjeet}[1]{\par\textbf{\color{orange}[Inderjeet: #1]}\par}
\else
  \newcommand{\Rami}[1]{\ignorespaces}
  \newcommand{\Aviram}[1]{\ignorespaces}
  \newcommand{\MergeNote}[1]{\ignorespaces}
  \newcommand{\Eyal}[1]{\ignorespaces}
  \newcommand{\Shaun}[1]{\ignorespaces}
  \newcommand{\Chen}[1]{\ignorespaces}
  \newcommand{\inderjeet}[1]{\ignorespaces}
\fi

\journal{Applied Soft Computing}

\begin{document}

\begin{frontmatter}

\title{\texorpdfstring{I$^2$C}{I2C}-GRPO: Insight Importance Checklist for Reinforcement Learning in Multi-Agent Network Alert Triage}

\author[bgu]{Eyal Zloof\texorpdfstring{\fnref{equal}}{}}
\author[bgu]{Shaun Shuster\texorpdfstring{\fnref{equal}}{}}
\author[jct,bgu]{Aviram Zilberman\texorpdfstring{\corref{cor1}}{}}
\author[bgu]{Chen Frydman}
\author[fujitsu]{Inderjeet Singh}
\author[fujitsu]{Andr\'{e}s Murillo}
\author[bgu]{Yosi Hadad}
\author[bgu]{Asaf Shabtai}
\author[bgu]{Yuval Elovici}
\author[bgu]{Rami Puzis}

\fntext[equal]{These authors contributed equally.}
\cortext[cor1]{Corresponding author.}

\affiliation[bgu]{organization={Ben-Gurion University of the Negev},
            city={Beer Sheva},
            country={Israel}}
\affiliation[jct]{organization={Jerusalem College of Technology},
            city={Jerusalem},
            country={Israel}}
\affiliation[fujitsu]{Fujitsu}

\begin{abstract}
Network alert labels indicate whether an alert is malicious and give little guidance on what evidence an agent should seek next. We propose \ac{I2C}-GRPO to train a Triage Agent on fixed investigation prefixes. Expert models propose actions, and a separate panel ranks the forensic insights extracted from those proposals. We score each candidate action by the weighted coverage of the forensic insights it targets, without executing the action. A rank exponent controls how strongly this additive fuzzy measure rewards the highest-ranked insights.

We train Qwen3-4B on DARPA 2000 and evaluate it on five alert sets from three benchmarks, covering DDoS, Ares botnet, and infiltration traffic. Across three training seeds, quadratic checklist weighting improves mean F1 by 3.1 to 5.8 percentage points over uniform weighting and by 4.9 to 9.7 points over DIVA. On CSE-CIC-IDS2018 DDoS, the proposed method attains 0.869 F1 and a 3.15\% false-positive rate, compared with 0.701 and 10.27\% for GPT-5-mini. On infiltration, the full workflow improves F1 from 0.515 with alert-only triage to 0.731.
\end{abstract}

\begin{highlights}
\item Ranked insight checklists reward intermediate choices without full rollouts
\item Rank weights improve F1 over uniform weights in all five evaluations
\item The DARPA-trained policy is evaluated on DDoS, botnet, and infiltration alerts
\item Full investigation improves infiltration F1 from 0.515 to 0.731
\end{highlights}


\end{frontmatter}

\acresetall


\begin{acronym}
  \acro{AI}{Artificial Intelligence}
  \acro{API}{Application Programming Interface}
  \acro{APT}{Advanced Persistent Threat}
  \acro{C2}{command-and-control}
  \acro{CoT}{Chain-of-Thought}
  \acro{CT}{Cyber Twin}
  \acro{CTI}{Cyber Threat Intelligence}
  \acro{DDoS}{Distributed Denial of Service}
  \acro{DT}{Digital Twin}
  \acro{DSR}{defense success rate}
  \acro{DPO}{direct preference optimization}
  \acro{EDR}{Endpoint Detection and Response}
  \acro{GNS3}{Graphical Network Simulator 3}
  \acro{ICS}{Industrial Control System}
  \acro{I2C}[I$^2$C]{Insight Importance Checklist}
  \acro{IDS}{Intrusion Detection System}
  \acro{IoC}{Indicators of Compromise}
  \acro{IM}{Incident Management}
  \acro{IR}{Incident Report}
  \acro{IPS}{Intrusion Prevention System}
  \acro{KG}{Knowledge Graph}
  \acro{LLM}{large language model}
  \acro{ME}{Mitigation Effectiveness}
  \acro{ML}{Machine Learning}
  \acro{NLP}{Natural Language Processing}
  \acro{OCE}{On Call Engineer}
  \acro{PE}{Prompt Engineering}
  \acro{RAG}{Retrieval-augmented generation}
  \acro{SIEM}{Security Information and Event Management}
  \acro{SOAR}{Security Orchestration Automation and Response}
  \acro{SOC}{Security Operations Center}
  \acro{TTP}{Tactics, Techniques, and Procedures}
  \acro{PT}{Penetration Testing}
  \acro{ASP}{Attack Success Parity}
  \acro{GRPO}{Group Relative Policy Optimization}
  \acro{SFT}{Supervised Fine-Tuning}
  \acro{RLVR}{Reinforcement Learning from Verifiable Rewards}
  \acro{POMDP}{Partially Observable Markov Decision Process}
  \acro{PCAP}{packet capture}
  \acro{PPO}{Proximal Policy Optimization}
  \acro{RL}{reinforcement learning}
  \acro{RLAIF}{reinforcement learning from AI feedback}
  \acro{dDPO}{distilled direct preference optimization}
  \acro{dRLAIF}{distilled reinforcement learning from AI feedback}
\end{acronym}

\section{Introduction}
\label{sec:introduction}

A burst of connections to one service could reflect ordinary demand or coordinated activity. To decide how to proceed, an analyst might ask whether the sources started sending traffic together or whether earlier alerts involve the same hosts. Choosing what evidence to seek is part of network alert triage, especially in security operations centers that handle large volumes of alerts requiring no action~\cite{alahmadi202299,yang2024true}.

A malicious or benign label can score the final verdict but gives little guidance on which questions to ask. Evaluating each candidate question by continuing the investigation would require repeated evidence retrieval and decisions by several agents. We propose \ac{I2C}-GRPO to train the Triage Agent to ask questions that target important forensic insights. Fixed helper agents gather the initial context, retrieve evidence in response to questions, and review proposed verdicts.

\begin{figure}[!htbp]
    \centering
    \includegraphics[width=\textwidth]{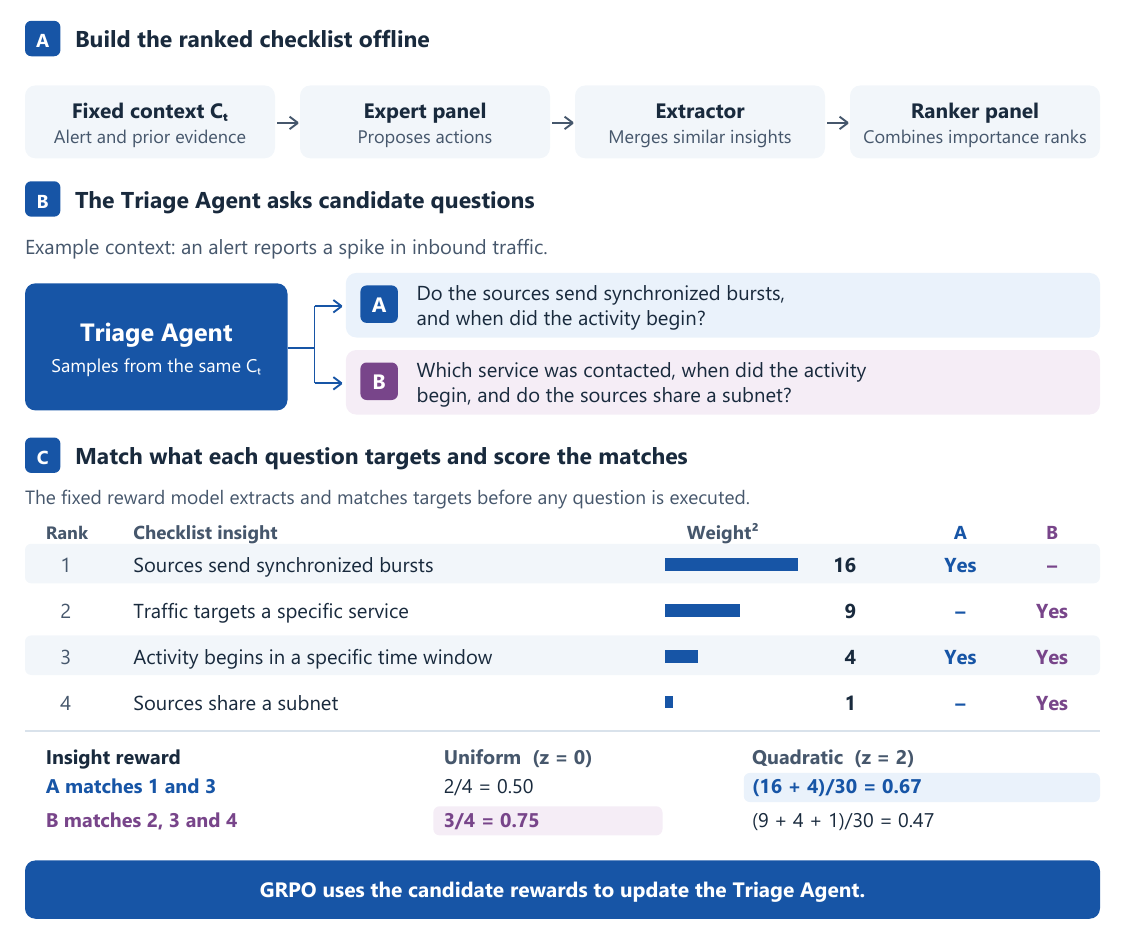}
    \caption{How \ac{I2C}-\acs{GRPO} turns expert rankings into a reward for the Triage Agent. \textbf{(A)} Expert proposals and importance rankings yield a checklist for a fixed investigation context. \textbf{(B)} The Triage Agent samples candidate questions from that context. \textbf{(C)} A fixed reward model matches the insights each question targets without executing the question. In this example, uniform weights favor question B, which matches three entries. Quadratic weights favor question A, which matches the highest-ranked entry and one other entry. These insight scores contribute to the policy update via Eq.~\eqref{eq:insight-reward}.}
    \label{fig:workflow}
\end{figure}

For each recorded investigation prefix, expert models propose possible next actions. An extractor identifies the forensic insights these actions target, and a separate panel orders the insights by importance using mean rank. During training, the Triage Agent samples candidate actions for the same prefix. A fixed matcher finds the checklist entries each action targets, and their weights determine the insight reward without executing the action. The rank exponent controls how much weight goes to the top of the checklist, from uniform through linear to quadratic weighting~\cite{stillwell1981comparison,barron1996decision}. Figure~\ref{fig:workflow} shows why a single high-ranked insight can outweigh several lower-ranked ones.

We train on DARPA 2000 LLDOS 1.0 and evaluate on five alert sets from DARPA 2000, CSE-CIC-IDS2018, and CICDDoS2019~\cite{darpa2000lldos,csecicids2018,cicddos2019}. Quadratic weighting improves mean F1 over uniform weighting and DIVA, which weights checklist entries by how well they distinguish candidate responses~\cite{cook2026check}. An ablation using GPT-5-mini for both the trajectories and the checklist proposals also improves on the GPT-5-mini Triage Agent across all four external test sets. On infiltration, the full workflow reaches 0.731 F1 versus 0.515 for one-step triage with the same initial evidence.

\subsection{Contributions}

\begin{enumerate}
    \item A method for training the Triage Agent on fixed investigation prefixes by scoring candidate actions against consensus-ranked forensic insights.
    \item An analysis of how the rank exponent changes orness and dispersion, with controlled comparisons of uniform, linear, and quadratic weighting.
    \item An evaluation across five alert sets that tests checklist weights and insight sources and compares full investigation with one-step triage.
\end{enumerate}

\FloatBarrier

\section{Background}
\label{sec:background}

\subsection{Cybersecurity Alert Triage}
\label{subsec:background-triage}

Intrusion detection systems raise alerts from signals such as packet patterns or flow statistics. Triage begins after an alert is raised and uses additional evidence to determine whether the activity is malicious~\cite{ahmed2016survey,sharafaldin2018toward}. Evidence can be distributed across packets, earlier alerts, and host records.

\subsection{Sequential Investigation and Process Rewards}
\label{subsec:background-sequential}

Alert triage is partially observable. The agent sees the initial alert and the evidence it has retrieved, and each question can yield a new observation. A terminal reward scores the verdict. A process reward scores intermediate actions, such as how well a question targets a forensic finding relevant to the case~\cite{nguyen2021deep}.

Our training procedure samples and scores candidate actions at fixed investigation prefixes. Because those actions do not advance the investigation during training, each prefix defines a one-step contextual bandit. At evaluation, the policy runs full investigations and can use evidence retrieved by its previous questions. Section~\ref{subsec:model} defines the policy's context and actions.

\subsection{Group Relative Policy Optimization}
\label{subsec:background-grpo}

\ac{PPO} estimates advantages with a learned value function~\cite{schulman2017ppo}. \ac{GRPO} estimates them from groups of candidates sampled for the same context~\cite{shao2024deepseekmath}. Each candidate's advantage depends on its reward relative to the group's mean and standard deviation (Eq.~\eqref{eq:grpo-advantage}). In our setting, each group contains possible next actions for the same investigative decision.

\section{Related Work}\label{sec:related-work}

\subsection{Network-Intrusion Benchmarks and Alert Triage}

DARPA 2000 provides multi-stage attack traces, and CIC benchmarks add labeled packet and flow data for a broader range of attacks~\cite{darpa2000lldos,hutchins2011killchain,sharafaldin2018toward,cicids2017,csecicids2018,cicddos2019}. Most detection methods use engineered features from flows or packets to detect malicious activity~\cite{sommer2010outside,buczak2015survey,ahmed2016survey,leevy2021detecting,ren2022id,agate2025mides,li2024chcg}. Triage addresses the alerts those detectors produce, many of which are false positives~\cite{zhang2022prehmm,gong2024alertpro}. CrossAlert, for example, classifies Snort alerts using features extracted from the traffic~\cite{niknami2024crossalert}. In our setting, the Triage Agent also chooses what evidence to retrieve before proposing its verdict.

\subsection{Reinforcement Learning for Cyber Defense}

Prior \ac{RL} work optimizes detector decisions, feature selection, alert prioritization, and adaptive response~\cite{nguyen2021deep,lopezmartin2020application,lopezmartin2021rbfnn,ren2022id,sethi2021attention,gong2024alertpro,jalalvand2025l2dhf}. It has also been applied to malware investigation and threat hunting~\cite{moser2025malwareinvestigation,ahmad2025securenetrl}. Recent combinations of \acp{LLM} and RL use a knowledge graph for mitigation selection, prioritize traffic before contextual analysis, or calibrate a cloud intrusion detector~\cite{towhid2025knowledgeinfused,sahay2026policyguided,ali2026cloudids}. The policy we train makes a different decision during the investigation: which forensic question to ask given the evidence so far.

\subsection{\ac{LLM} Agents for Cyber Investigation}

HuntGPT and IDS-Agent use language models for threat-hunting explanations and explainable detection~\cite{ali2026huntgpt,li2024idsagent}. CORTEX is closest to our architecture. Its prompted agents collaborate to analyze alerts and acquire evidence, and its authors release analyst investigation traces from a production security operations center~\cite{wei2025cortex}. Our workflow likewise assigns different roles to agents and optimizes the Triage Agent within the fixed team~\cite{stone2010ad}. Holmes investigates DDoS from packet-derived evidence packs, and CFA-Bench evaluates forensic agents on service, packet, and log evidence~\cite{chen2026holmes,de2025cfa}.

Other systems use retrieval for packet analysis, expose executable investigation environments, support incident response, or focus on DDoS, unknown attacks, and multi-stage detection~\cite{briliyant2025network,excitinbench2025,barnes2026opensec,chona2026cyber,gao2026incontextir,cui2025aglhunter,chougle2025ddosllm,alsuwaiket2025zerodayllm,ali2026unknownattack,li2025madllm}. Surveys discuss the relevance of tool use and the risks of unreliable model judgments~\cite{zhang2025llms,llmtaxonomy2025}. Outside security, expected uncertainty reduction provides a model-based criterion for selecting questions~\cite{hu2024uncertainty}. Our checklist provides a proxy based on the importance of the forensic insights a question targets.

\subsection{Process Rewards and Checklist Feedback}

Model judges and rubric-based rewards supply training feedback where verifiable labels are sparse~\cite{zheng2023judging,gu2024survey,Yuan2024SelfRewarding,lee2024rlaif,gunjal2026rubrics}. Step-level supervision has been studied in reasoning and code security~\cite{lightman2024let,yu2026seccodeprm}. Optimizing a proxy reward can still degrade the intended objective~\cite{gao2023scaling}. Using several judges can reduce reliance on one model's judgments~\cite{verga2024replacing}.

RLCF builds instruction-specific checklists and aggregates per-item scores into a response reward~\cite{viswanathan2025checklistsbetterrewardmodels}. CM2 mines binary per-turn criteria from a reference tool-use trajectory and assigns rewards at selected points~\cite{zhang2026cm2}. DIVA weights checklist items by how much their scores vary across the candidate group~\cite{cook2026check}. In \ac{I2C}-GRPO, a separate ranker panel orders the insights for each context before training. The reward uses that fixed order to weight matches for each candidate action. Table~\ref{tab:positioning} summarizes these comparisons.

\begin{table*}[t]
\centering
\caption{Positioning against the closest checklist-reward and agentic-triage methods.}
\label{tab:positioning}
\scriptsize
\begin{tabularx}{\textwidth}{@{}>{\raggedright\arraybackslash}p{1.6cm}>{\raggedright\arraybackslash}X>{\raggedright\arraybackslash}X>{\raggedright\arraybackslash}X>{\raggedright\arraybackslash}p{1.5cm}>{\raggedright\arraybackslash}p{1.7cm}@{}}
\toprule
Method & Criteria source & Item weighting & Reward granularity & Trained agent & Domain \\
\midrule
RLCF~\cite{viswanathan2025checklistsbetterrewardmodels} & Instruction-specific checklist generated per prompt & Per-item scores aggregated into one reward & Whole response & Yes & Instruction following \\
CM2~\cite{zhang2026cm2} & Per-turn criteria mined from one reference trajectory & Binary pass/fail criteria & Trajectory, turn, or step (sparse) & Yes & Agentic tool use \\
DIVA~\cite{cook2026check} & Checklist items & Within-group item variance & Whole response & Yes & Checklist-based RL \\
CORTEX~\cite{wei2025cortex} & -- (no reward) & -- & -- & No (prompted) & SOC alert triage \\
\textbf{\ac{I2C}-GRPO} & Forensic insights proposed by an expert panel and ordered by a ranker panel & Consensus rank exponent ($z=2$) & Candidate action at a fixed prefix & Yes (one agent in a fixed team) & Network alert triage \\
\bottomrule
\end{tabularx}
\end{table*}

\subsection{Consensus and Rank Aggregation under Uncertainty}
\label{subsec:rw-soft}

Group decision methods combine rankings or fuzzy preferences and can quantify agreement using Kendall's $W$~\cite{herrera2002consensus,kendall1939problem}. Borda scoring combines rank positions, while Kemeny aggregation seeks a ranking that minimizes disagreement with the input rankings~\cite{kemeny1959mathematics,dwork2001rank}. Rank-exponent weights map ordinal positions to criterion weights~\cite{stillwell1981comparison,barron1996decision}. Ordered weighted averaging, fuzzy sets, and fuzzy measures provide related tools for weighting and aggregating judgments~\cite{yager1988ordered,zadeh1965fuzzy,grabisch1996application,blanco2019bibliometric}. In our method, the panel's mean rank determines the order, and the resulting weights define an additive fuzzy measure on the set of matched insights.

Recent soft-computing work uses \acp{LLM} to build adaptive agents or design rewards for a scheduler, and fuzzy or related methods have been used to address class imbalance in intrusion detection~\cite{zhang2025simulating,hong2025large,arafah2025anomaly}. Here, the aggregated ranking provides the process reward for a language model's investigative actions. Our evaluation compares this reward with checklist baselines and varies the ordering, weights, and expert sources within the same triage workflow.

\section{Investigation Scope}
\label{sec:threat-model}

We evaluate post-mortem triage of network intrusion alerts with access to recorded traffic and associated metadata (Section~\ref{subsec:framework}). The evaluation covers the sampled benchmark captures described in Section~\ref{subsec:eval-protocol}. Real-time mitigation and adversarial manipulation of the monitoring or agent infrastructure are outside its scope.

\section{System Model and Methodology}
\label{sec:method}

\subsection{Investigation Context and Policy}
\label{subsec:model}

At step $t$, the Triage Agent observes a context $C_t=(o_0,a_0,\ldots,a_{t-1},o_t)$ of earlier actions and observable evidence. The context is the serialized conversation, beginning with the alert and including evidence retrieved by the helper agents. The policy $\pi_\theta(a_t\mid C_t)$ can ask a question or propose a verdict. Each question can reveal an observation, so evaluated investigations are sequential and partially observable. Training samples candidates on fixed prefixes and scores them without executing their actions (Section~\ref{subsec:grpo}).

\subsection{Agent Roles and Evidence}
\label{subsec:framework}

\begin{figure}[!htbp]
    \centering
    \includegraphics[width=\linewidth]{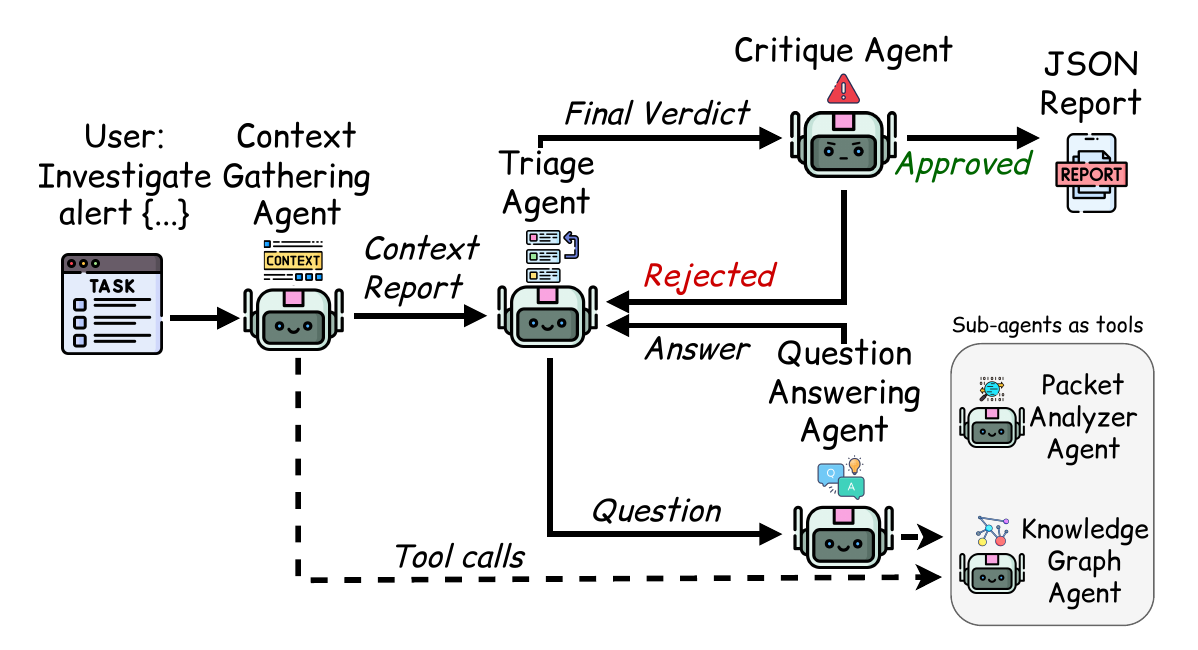}
    \caption{The Triage Agent asks follow-up questions and proposes verdicts. The Critique Agent can accept a verdict or return it for further investigation.}
    \label{fig:agent-workflow}
\end{figure}

Only the Triage Agent is optimized. Three fixed helper agents prepare evidence, answer questions, and review verdicts (Figure~\ref{fig:agent-workflow}).

\begin{itemize}
    \item \textbf{Context Gathering Agent:} Builds the initial context from the alert, packet-derived evidence, alert history, and host inventory.
    \item \textbf{Triage Agent:} Asks a forensic question or proposes a verdict with a classification, confidence, and impact score.
    \item \textbf{Question Answering Agent:} Retrieves evidence from indexed packet captures and the \ac{KG} in response to the Triage Agent's question. Its answer becomes part of the next context.
    \item \textbf{Critique Agent:} Checks whether the evidence supports the proposed classification, confidence, and reasoning. If it rejects the verdict, the investigation continues.
\end{itemize}

\paragraph{Knowledge Graph Evidence}
The \ac{KG} is a read-only Neo4j evidence store used by the Context Gathering and Question Answering Agents for structured factual context. It represents alerts, IP addresses, detection rules, and network segments as nodes, linking alerts to endpoints and rules and linking hosts to network segments. The agents use constrained Cypher queries to retrieve host inventory, network topology, rule, and alert-history facts; packet- and stream-level questions are instead answered from the indexed raw-PCAP database. Graph queries cannot write data or access evaluation labels, verdicts, or benchmark-role metadata.

\subsection{Offline \ac{I2C} Construction}
\label{subsec:candidate}

\subsubsection{Expert Action Generation}

We recorded investigations on the DARPA 2000 training alerts using GPT-5-mini as the Triage Agent. We truncate each investigation immediately before each Triage Agent action to obtain the training prefixes $C_t\in\mathcal{H}$. These contexts are off-policy for Qwen3-4B, whose actions can produce different contexts at evaluation.

The main checklist configuration uses Gemini 3.1 Pro, GPT-5.5, and Claude Opus 4.7 to propose next actions for each prefix~\cite{google2026gemini31pro,openai2026gpt55,anthropic2026opus47}.
For each prefix, we sample $N$ actions from each expert policy:

\[
a_{t,e}^{(i)} \sim \pi_e(\cdot \mid C_t), \qquad i \in \{1,\dots,N\}, \quad e \in \mathcal{E}
\]

Here $\mathcal{E}$ is the expert set and $\pi_e$ is the fixed policy of expert $e$.

Each expert action may correspond to a natural-language forensic question or a decision to conclude the investigation. For example, given evidence of suspicious traffic, the generated action set may contain:

\begin{itemize}
    \item ``Was the vulnerable service exploited?''
    \item ``Which hosts communicated with the suspicious host?''
    \item ``What was the timestamp of the alert?''
\end{itemize}

\subsubsection{Insight Extraction and \ac{I2C} Construction}

A \emph{forensic insight} is an atomic, declarative finding, such as a host having communicated with a known compromised host. The action is the question or verdict the Triage Agent produces. For example, a question about communications can target the finding that two hosts exchanged traffic. The checklist ranks such findings, and the training reward credits actions that target them. This provides a proxy for investigative quality that can be scored immediately. The evaluation measures triage outcomes, while individual insights lack ground-truth labels (Section~\ref{subsec:limitations}).

Given the current Triage Agent context $C_t$, we aggregate the candidate actions produced by all frontier models:

\[
\mathcal{A}_t^{\mathrm{expert}} = \Big\{ a_{t,e}^{(i)} \;\Big|\; e \in \mathcal{E}, \, i \in \{1,\dots,N\} \Big\}
\]

The \emph{extractor} turns this pool of actions into atomic insights. It merges equivalent insights and records the source expert text for each one. A separate panel of three rankers orders the insights to form the checklist $Z_t$.

The offline extractor is DeepSeek-V4-Pro~\cite{deepseek2026v4}, and the ranker panel consists of DeepSeek-V4-Pro, Kimi K2.7 Code, and Qwen3.7-Max. During training, Gemma 4 31B IT extracts and matches the insights targeted by each candidate (Section~\ref{sec:i2c-guided-insight-reward}). The three helper agents use Qwen3.5-35B-A3B. Table~\ref{tab:model-versions} gives the API identifiers and checkpoints. Hosted models used the provider revisions available at run time and the provider-default sampling settings.

\begin{table}[t]
\centering
\caption{Models used in this work, with hosted API identifiers and local checkpoint names. DeepSeek-V4-Pro was accessed through Microsoft Azure.}
\label{tab:model-versions}
\scriptsize
\begin{tabularx}{\textwidth}{@{}p{2.3cm}p{2.6cm}X@{}}
\toprule
Model & Role & API identifier or checkpoint \\
\midrule
Gemini 3.1 Pro & Expert panel & \nolinkurl{gemini-3.1-pro-preview} \\
GPT-5.5 & Expert panel & \nolinkurl{gpt-5.5} \\
Claude Opus 4.7 & Expert panel & \nolinkurl{claude-opus-4-7} \\
DeepSeek-V4-Pro & Extractor; ranker & \nolinkurl{DeepSeek-V4-Pro} (Azure) \\
Kimi K2.7 Code & Ranker & \nolinkurl{kimi-k2.7-code} \\
Qwen3.7-Max & Ranker & \nolinkurl{qwen3.7-max} \\
GPT-5-mini & Trajectory generator, baseline, insight source & \nolinkurl{gpt-5-mini} \\
Gemma 4 31B IT & Online reward model & \nolinkurl{google/gemma-4-31B-it} \\
Qwen3.5-35B-A3B & Helper agents & \nolinkurl{Qwen/Qwen3.5-35B-A3B} \\
Qwen3-4B & Trained Triage Agent base & \nolinkurl{Qwen/Qwen3-4B-Instruct-2507} \\
\bottomrule
\end{tabularx}
\end{table}

\begin{equation}
\label{eq:i2c}
Z_t = \big( z_t^{(1)}, z_t^{(2)}, \dots, z_t^{(K)} \big)
\end{equation}
where $z_t^{(k)}$ is the insight at position $k$ in descending order of panel-assessed importance for $C_t$. Checklist length $K=K_t$ can vary across contexts.

\paragraph{Rank aggregation and ties}\label{subsubsec:aggregation}
Let $\rho_r(k)\in\{1,\dots,K\}$ be the position that ranker $r\in\mathcal{J}$ assigns to insight $k$. We order insights by increasing mean rank,
\[
\bar{\rho}(k)=\frac{1}{|\mathcal{J}|}\sum_{r\in\mathcal{J}}\rho_r(k).
\]
This is equivalent to Borda ordering because every ranker orders the same complete set and the Borda score is $|\mathcal{J}|\,(K-\bar{\rho}(k))$. The mean is the equally weighted case of ordered weighted averaging~\cite{yager1988ordered}. We break ties uniformly at random and hold each resulting checklist fixed throughout training.

\subsection{\ac{I2C}-Guided Insight Reward Design}

\label{subsec:reward}

The reward scores weighted checklist coverage at each training prefix. It assigns no explicit cost to extra questions or to either type of classification error. We assess the classification tradeoff using precision and recall.

\subsubsection{Rank-Exponent Insight Weights}
\label{subsubsec:ranking}
To account for the varying importance of insights in $Z_t$, we assign each insight a rank-based weight according to its position in the aggregated \ac{I2C}:

\begin{equation}
\label{eq:rank}
w_k = (K + 1) - k
\end{equation}
where $k \in \{1, \dots, K\}$ denotes the position of insight $z_t^{(k)}$ in the checklist.
The reward uses the powers $w_k^z$ with an \emph{importance-concentration exponent} $z\ge0$. These are the rank-exponent weights of multi-attribute decision analysis~\cite{stillwell1981comparison}, a family that also contains equal weights ($z=0$) and rank-sum weights ($z=1$) and is closely related to other rank-based weight approximations such as rank-order centroid weights~\cite{barron1996decision}. Larger $z$ concentrates the weight on the top of the consensus ranking: for a checklist of $K=10$ insights, the three top-ranked insights carry 30\% of the total weight at $z=0$, 49\% at $z=1$, and 64\% at $z=2$; Section~\ref{subsubsec:soft-interpretation} characterizes this concentration through the orness of the weights. The proposed method uses $z=2$; Section~\ref{subsec:exponent} compares $z\in\{0,1,2\}$, and Section~\ref{subsec:reward-ablation} tests the checklist order.

\subsubsection{\ac{I2C}-Guided Insight Reward}\label{sec:i2c-guided-insight-reward}

For candidate $i$, Gemma 4 31B IT uses $C_t$ and the candidate's output to extract insights from the reasoning and the proposed question or verdict. The model then matches the resulting set $\widehat{\mathcal{I}}_t^{(i)}$ to the fixed checklist $Z_t$. Over $G$ candidates, this yields a binary \emph{insight-capture matrix}
\begin{equation}
\begin{aligned}
M_t &=[M_{t,i,k}]\in\{0,1\}^{G\times K},\\
M_{t,i,k}
&=\mathbb{1}\!\left[
\substack{
\exists\,\hat{z}\in\widehat{\mathcal{I}}_t^{(i)}:\\
\hat{z}\text{ matches }z_t^{(k)}
}
\right].
\end{aligned}
\label{eq:capture-matrix}
\end{equation}
Let $H_t^{(i)}=\{k:M_{t,i,k}=1\}$ be the entries matched by candidate $i$. Their normalized weights determine the insight reward:
\begin{equation}
\label{eq:insight-reward}
r_{t}^{\text{insight},(i)} =
\frac{\sum_{k=1}^{K} M_{t,i,k}\,w_k^{z}}
{\sum_{k=1}^{K}w_k^{z}}, \qquad z=2.
\end{equation}

Figure~\ref{fig:workflow} illustrates how the weights change the relative scores of two candidates. The reward measures the relevance of the insights a candidate targets at $C_t$. Full-workflow evaluation measures the resulting triage performance.

\subsubsection{Soft-Computing Interpretation}
\label{subsubsec:soft-interpretation}

\begin{definition}[Rank-exponent capacity]
\label{def:capacity}
For a checklist $Z_t$ with $K$ entries and an exponent $z\ge0$, let $p_k=w_k^{z}/\sum_{j=1}^{K}w_j^{z}$ and define
\begin{equation}
\label{eq:fuzzy-measure}
\mu_t(S)=\sum_{k\in S}p_k, \qquad S\subseteq\{1,\dots,K\}.
\end{equation}
\end{definition}
The set function $\mu_t$ satisfies $\mu_t(\varnothing)=0$, $\mu_t(\{1,\dots,K\})=1$, and $S\subseteq T\Rightarrow\mu_t(S)\le\mu_t(T)$, so it is a normalized fuzzy measure (capacity) on the checklist, and the insight reward is its value on the captured-index set, $r_t^{\text{insight},(i)}=\mu_t\big(H_t^{(i)}\big)$.

\begin{remark}
\label{rem:additive}
$\mu_t$ is additive, so it is fully determined by the $K$ singleton weights $p_k$, and Eq.~\eqref{eq:insight-reward} is a normalized weighted sum. The capture vector $x_k=M_{t,i,k}$ is binary, and for binary inputs the discrete Choquet integral with respect to \emph{any} capacity $\nu$ equals $\nu(\{k:x_k=1\})$~\cite{grabisch1996application}. The soft-computing analysis therefore concerns how consensus ranks determine the weights and how strongly those weights concentrate on the top of the checklist.
\end{remark}

Because the weights depend only on consensus position, $\mathbf{p}=(p_1,\dots,p_K)$ is an OWA-type weighting vector over positions ordered from most to least important~\cite{yager1988ordered}. Yager's orness and dispersion of such a vector are
\begin{equation}
\label{eq:orness}
\operatorname{orness}(\mathbf{p})=\sum_{k=1}^{K}\frac{K-k}{K-1}\,p_k, \qquad
\operatorname{disp}(\mathbf{p})=-\sum_{k=1}^{K}p_k\ln p_k .
\end{equation}
Orness is $0.5$ for equal weights and approaches $1$ as all weight moves to the top-ranked insight; dispersion is the entropy of the weights.

\begin{proposition}
\label{prop:orness}
For fixed $K\ge2$, $\operatorname{orness}(\mathbf{p})$ is strictly increasing and $\operatorname{disp}(\mathbf{p})$ is strictly decreasing in $z\ge0$, with $\operatorname{orness}=0.5$ and $\operatorname{disp}=\ln K$ at $z=0$.
\end{proposition}
\begin{proof}
Write $j=K{+}1{-}k\in\{1,\dots,K\}$, so $p_j\propto j^{z}=e^{z\ln j}$ is an exponential family with sufficient statistic $\ln j$, and $\operatorname{orness}=\mathbb{E}_z[(j-1)/(K-1)]$. For $z'>z$, the ratio $p^{(z')}_j/p^{(z)}_j\propto j^{z'-z}$ is strictly increasing in $j$, so $\mathbf{p}^{(z')}$ dominates $\mathbf{p}^{(z)}$ in the monotone likelihood ratio order and hence first-order stochastically; since $(j-1)/(K-1)$ is strictly increasing and $\mathbf{p}^{(z')}\neq\mathbf{p}^{(z)}$, the expectation strictly increases. For dispersion, $\ln p_j=z\ln j-\ln\Lambda(z)$ with $\Lambda(z)=\sum_j j^{z}$, so $\operatorname{disp}=\ln\Lambda(z)-z\,\mathbb{E}_z[\ln j]$ and $\frac{d}{dz}\operatorname{disp}=\mathbb{E}_z[\ln j]-\mathbb{E}_z[\ln j]-z\operatorname{Var}_z(\ln j)=-z\operatorname{Var}_z(\ln j)$, which is negative for $z>0$ because $\ln j$ is not constant for $K\ge2$. The values at $z=0$ follow from $p_j=1/K$.
\end{proof}

Table~\ref{tab:orness} gives an example for ten insights. Section~\ref{subsec:exponent} relates the change in orness to the observed error rates.

\begin{table}[t]
\centering
\caption{Orness and dispersion (Eq.~\eqref{eq:orness}) of the rank-exponent weighting vector for a checklist of $K=10$ insights, with the share of the total weight on the top-ranked insight and on the three top-ranked insights. Values are computed analytically, not measured.}
\label{tab:orness}
\scriptsize
\begin{tabular}{@{}lccccc@{}}
\toprule
$z$ & Scheme & Orness & Dispersion (nats) & Top-1 weight & Top-3 weight \\
\midrule
0 & Uniform & 0.500 & 2.303 & 0.100 & 0.300 \\
1 & Rank sum & 0.667 & 2.151 & 0.182 & 0.491 \\
2 & Proposed & 0.762 & 1.921 & 0.260 & 0.636 \\
3 & Not evaluated & 0.819 & 1.719 & 0.331 & 0.741 \\
\bottomrule
\end{tabular}
\end{table}

Graded matches could give partial credit through a Choquet integral. A non-additive capacity could account for redundancy or synergy among insights, and weights informed by ranker disagreement could reflect uncertainty in their importance.

\subsubsection{Auxiliary Rewards}

The two auxiliary terms award the following bonuses when their conditions are met and zero otherwise. We did not tune their magnitudes in a sensitivity sweep.

\begin{itemize}
    \item \textbf{Format reward} ($0.05$). The judge checks that the output contains reasoning followed by either a question or a verdict with classification, confidence, and impact scores.
    \item \textbf{Question reward} ($0.2$). The judge checks for an investigative question. This larger bonus encourages continued evidence gathering.
\end{itemize}

\paragraph{Termination and the final verdict}
Training assigns a bonus to questions and can also reward checklist insights expressed in a terminating action. The reward excludes verdict correctness and supplies no explicit stopping objective. In full-workflow evaluation, the Critique Agent must accept the Triage Agent's verdict for the investigation to end. Investigations have no fixed step cap.

\subsubsection{Total Reward}

The three reward functions return already-scaled values, and GRPO sums them with unit aggregation weights:

\begin{equation}
\label{eq:total-reward}
r_t^{(i)} =
r_t^{\text{insight},(i)}
+r_t^{\text{format},(i)}
+r_t^{\text{question},(i)}.
\end{equation}

\subsection{Policy Optimization with \ac{GRPO} over Offline Contexts}
\label{subsec:grpo}

At each fixed prefix $C_t$, the current Triage Agent samples $G$ candidate next actions on-policy. Each candidate is scored on the same context and checklist without running the investigation forward.

Given rewards $r_t^{(1)},\dots,r_t^{(G)}$ for the Triage Agent actions sampled at context $C_t$, the relative advantage of candidate $i$ is
\begin{equation}
\label{eq:grpo-advantage}
A_{t,i}
=
\frac{
r_t^{(i)} - \frac{1}{G}\sum_{j=1}^{G} r_t^{(j)}
}{
\text{std}(r_t^{(1)},\dots,r_t^{(G)})+\varepsilon
}
\end{equation}
where $\varepsilon$ is a small constant for numerical stability. When all $G$ candidates receive the same reward, their policy advantages are zero.

The standard per-sequence \ac{GRPO} surrogate illustrates the role of clipping~\cite{shao2024deepseekmath}:
\begin{equation}
\label{eq:grpo-objective}
\mathcal{L}(\theta)=\frac{1}{G}\sum_{i=1}^{G}\frac{1}{|a_t^{(i)}|}\sum_{\ell\in\mathcal{T}_i}\min\!\Big(\kappa_{i,\ell}A_{t,i},\;\mathrm{clip}(\kappa_{i,\ell},1{-}\epsilon,1{+}\epsilon)A_{t,i}\Big)-\beta\,D_{\mathrm{KL}}\big(\pi_\theta\,\Vert\,\pi_{\mathrm{ref}}\big),
\end{equation}
where $\kappa_{i,\ell}=\pi_\theta(a_{t,\ell}^{(i)}\mid C_t,a_{t,<\ell}^{(i)})/\pi_{\theta_{\mathrm{old}}}(a_{t,\ell}^{(i)}\mid C_t,a_{t,<\ell}^{(i)})$ is the token-level probability ratio. Here $\epsilon$ is the clip range, $\beta$ is the KL coefficient, and $\pi_{\mathrm{ref}}$ is the reference policy. The reported configuration sets $\beta=0$, which disables the KL term (Table~\ref{tab:hyperparameters}). Following Wang et al.~\cite{wang2026beyond}, the policy-gradient term is restricted to the set $\mathcal{T}_i$ of high-entropy tokens: the policy loss uses token positions in the highest-entropy 20\% (Table~\ref{tab:hyperparameters}). This token restriction was not ablated.

\subsection{Training Procedure}
\label{subsec:training}

Algorithm~\ref{alg:offline-i2c} constructs $\mathcal{D}=\{(C_t,Z_t):C_t\in\mathcal{H}\}$ by extracting and ranking insights for each recorded prefix. Algorithm~\ref{alg:online-grpo} uses those fixed checklists to score new candidates and update the Triage Agent.

\begin{algorithm}[t]
\caption{Offline \ac{I2C} construction}
\label{alg:offline-i2c}
\begin{algorithmic}[1]
\REQUIRE Truncated investigation histories $\mathcal{H}$, expert model set $\mathcal{E}$, ranker panel $\mathcal{J}$, and expert sample count $N$
\ENSURE Offline dataset $\mathcal{D} = \{(C_t, Z_t) : C_t\in\mathcal{H}\}$
\STATE $\mathcal{D} \gets \varnothing$
\FOR{each $C_t \in \mathcal{H}$}
    \STATE $\mathcal{A}_t^{\mathrm{expert}} \gets \varnothing$
    \FOR{each $e \in \mathcal{E}$}
        \FOR{$i = 1$ to $N$}
            \STATE Sample $a_{t,e}^{(i)} \sim \pi_e(\cdot \mid C_t)$
            \STATE $\mathcal{A}_t^{\mathrm{expert}} \gets \mathcal{A}_t^{\mathrm{expert}} \cup \{a_{t,e}^{(i)}\}$
        \ENDFOR
    \ENDFOR
    \STATE $\mathcal{I}_t^{\mathrm{expert}} \gets \mathrm{ExtractInsights}(\mathcal{A}_t^{\mathrm{expert}}, C_t)$
    \STATE $\mathcal{P}_t \gets \{\mathrm{RankInsights}_r(\mathcal{I}_t^{\mathrm{expert}}, C_t): r\in\mathcal{J}\}$
    \STATE $Z_t \gets \mathrm{AggregateRanks}(\mathcal{P}_t)$, resolving rank ties at random
    \STATE $\mathcal{D} \gets \mathcal{D} \cup \{(C_t, Z_t)\}$
\ENDFOR
\RETURN $\mathcal{D}$
\end{algorithmic}
\end{algorithm}

\begin{algorithm}[t]
\caption{GRPO training of the Triage Agent over offline contexts}
\label{alg:online-grpo}
\begin{algorithmic}[1]
\REQUIRE Offline dataset $\mathcal{D}$, policy $\pi_\theta$, number of epochs $E$, batch size $B$, and candidate count $G$
\ENSURE Optimized policy $\pi_\theta$
\STATE Initialize the GRPO optimizer and hyperparameters
\FOR{$e = 1$ to $E$}
    \STATE Sample a batch $\mathcal{B}$ of size $B$ from $\mathcal{D}$
    \FOR{each $(C_t, Z_t) \in \mathcal{B}$}
        \STATE Sample $\{a_t^{(i)}\}_{i=1}^{G} \sim \pi_\theta(\cdot \mid C_t)$
        \FOR{$i = 1$ to $G$}
            \STATE $\widehat{\mathcal{I}}_t^{(i)} \gets \mathrm{ExtractInsights}(a_t^{(i)}, C_t)$
            \STATE $M_{t,i,:} \gets \mathrm{MatchInsights}(\widehat{\mathcal{I}}_t^{(i)}, Z_t)$
            \STATE $r_t^{\text{insight},(i)} \gets \mathrm{InsightReward}(M_{t,i,:}, Z_t)$
            \STATE $r_t^{\text{format},(i)} \gets 0.05\,\mathbb{1}[\mathrm{FormatJudge}(a_t^{(i)})]$
            \STATE $r_t^{\text{question},(i)} \gets 0.2\,\mathbb{1}[\mathrm{QuestionJudge}(a_t^{(i)})]$
            \STATE $r_t^{(i)} \gets \mathrm{TotalReward}(r_t^{\text{insight},(i)}, r_t^{\text{format},(i)}, r_t^{\text{question},(i)})$
        \ENDFOR
        \STATE Update $\theta$ with GRPO using $\{(a_t^{(i)}, r_t^{(i)})\}_{i=1}^{G}$ and context $C_t$
    \ENDFOR
\ENDFOR
\RETURN $\pi_\theta$
\end{algorithmic}
\end{algorithm}

\FloatBarrier

\section{Evaluation}
\label{sec:evaluation}

\subsection{Training Setup for Compared Methods}
\label{subsec:eval-setup}

All trained variants start from Qwen3-4B and use the same truncated DARPA histories $\mathcal{H}$ (Section~\ref{subsec:candidate}). They differ in their training signal (Table~\ref{tab:design-decomposition}). The three helper agents use Qwen3.5-35B-A3B in every full-workflow evaluation.

\ac{SFT} imitates expert actions paired with their contexts. \ac{dDPO} learns from judge-labeled preferences between expert actions~\cite{tunstall2023zephyr,rafailov2023direct}. RLCF scores instruction-specific checklist items~\cite{viswanathan2025checklistsbetterrewardmodels}, and \ac{dRLAIF} uses a judge's direct score for each action~\cite{lee2024rlaif}. DIVA follows Cook et al.~\cite{cook2026check}, including their recommended matcher. Uniform, linear, and quadratic weighting use the same forensic checklist and Gemma 4 31B IT matcher. Gemma also labels preferences for dDPO and provides judge-based rewards for RLCF, dRLAIF, and the other checklist ablations.

GRPO variants use the Hugging Face TRL trainer. All trained variants use one epoch and a learning rate of $10^{-5}$. Table~\ref{tab:hyperparameters} gives the proposed method's training configuration. Training used two NVIDIA RTX 6000 Ada Generation GPUs, one for optimization and one for the reward model.

Full-workflow variants use the same Triage Agent prompts. At evaluation, trained Qwen3-4B variants use the recommended settings for Qwen3-4B-Instruct-2507: temperature $0.7$, top-$p$ $0.8$, top-$k$ $20$, and min-$p$ $0$. The checkpoint runs in non-thinking mode.\footnote{\url{https://huggingface.co/Qwen/Qwen3-4B-Instruct-2507}} GPT-5-mini uses the provider default for reasoning effort.

\begin{table*}[t]
\centering
\caption{Design choices of the trained Triage Agent variants. All start from Qwen3-4B and use the same training contexts $\mathcal{H}$.}
\label{tab:design-decomposition}
\scriptsize
\begin{tabularx}{\textwidth}{@{}>{\raggedright\arraybackslash}p{2.1cm}>{\raggedright\arraybackslash}X>{\raggedright\arraybackslash}p{2.6cm}>{\raggedright\arraybackslash}X>{\raggedright\arraybackslash}p{2.1cm}@{}}
\toprule
Variant & Supervision source & Item weighting & Scoring unit & Optimizer \\
\midrule
SFT & Expert-panel actions (imitation) & -- & Token likelihood of the expert action & Supervised \\
dDPO & Judge preferences between expert actions & -- & Chosen--rejected pair & DPO \\
dRLAIF & Judge score per sampled action & -- & Absolute scalar score & GRPO \\
RLCF & Instruction-specific checklist per $C_t$ & As in RLCF~\cite{viswanathan2025checklistsbetterrewardmodels} & Per-item judge score & GRPO \\
DIVA & Checklist feedback & Within-group item variance & Item scores as in Cook et al.~\cite{cook2026check} & GRPO \\
Uniform & \ac{I2C} items & Equal ($z=0$) & Extract-then-match (binary) & GRPO \\
Linear & \ac{I2C} items & Rank sum ($z=1$) & Extract-then-match (binary) & GRPO \\
\ac{I2C}-GRPO & \ac{I2C} items & Rank exponent ($z=2$) & Extract-then-match (binary) & GRPO \\
\bottomrule
\end{tabularx}
\end{table*}

\begin{table}[htbp]
\centering
\caption{Training hyperparameters for the proposed \ac{I2C}-GRPO variant.}
\label{tab:hyperparameters}
\begin{tabular}{ll}
\hline
Parameter & Value \\
\hline
Learning rate & 1e-5 \\
Per-device training batch size & 4 \\
Gradient accumulation steps & 4 \\
RL algorithm & GRPO \\
Number of generations & 8 \\
Training epochs & 1 \\
Reported checkpoint & End of epoch 1 \\
Optimizer & 8-bit AdamW \\
Scheduler & Cosine annealing \\
Warmup steps & 100 \\
Context window & 11,000 \\
Max completion length & 1024 \\
Temperature / top-$p$ & 1.0 / 1.0 \\
Top-entropy quantile~\cite{wang2026beyond} & 0.2 \\
KL coefficient $\beta$ & $0.0$ (KL penalty disabled) \\
Clipping coefficient $\epsilon$ & $0.2$ (ratio bounds $[0.8, 1.2]$) \\
Fine-tuning mode & LoRA ($r=16$, $\alpha=32$) \\
\hline
\end{tabular}
\end{table}

\subsection{Evaluation Protocol}
\label{subsec:eval-protocol}

We evaluate binary alert triage on five sets from three benchmarks (Table~\ref{tab:eval-sets}). Training uses 1,536 decision-point prefixes from DARPA 2000 LLDOS 1.0. The DARPA evaluation combines held-out alerts from LLDOS 1.0 and LLDOS 2.0.2, using the inside and DMZ captures. The 30 positive alerts cover five stages of the campaigns, with 18 from LLDOS 1.0 and 12 from LLDOS 2.0.2. The 120 benign alerts come from non-attack intervals, 60 from each sensor location. Ground truth is derived from the MIT Lincoln Laboratory scenario timelines and endpoint manifests. The alert-level partition precedes the extraction of training prefixes, and IP identifiers in the training split were randomized while preserving topology and timestamps.

\begin{table}[t]
\centering
\caption{Evaluation alert counts. Ares and infiltration are separate capture days within CSE-CIC-IDS2018.}
\label{tab:eval-sets}
\small
\begin{tabular}{@{}llrr@{}}
\toprule
Benchmark & Scenario & Malicious & Benign \\
\midrule
DARPA 2000~\cite{darpa2000lldos} & LLDOS 1.0 and 2.0.2 & 30 & 120 \\
CSE-CIC-IDS2018~\cite{csecicids2018} & LOIC and HOIC (20--21 Feb.) & 400 & 1,600 \\
CICDDoS2019~\cite{cicddos2019} & DDoS (11 Mar.) & 400 & 1,600 \\
CSE-CIC-IDS2018 & Ares botnet (2 Mar.) & 250 & 1,000 \\
CSE-CIC-IDS2018 & Infiltration (28 Feb.) & 200 & 800 \\
\bottomrule
\end{tabular}
\end{table}

The CSE-CIC-IDS2018 DDoS set contains 160 LOIC-HTTP, 160 HOIC, and 80 LOIC-UDP positives, plus 800 benign alerts from each of the February 20 and 21 captures. The Ares set uses March 2, and infiltration uses February 28. All three sets use the corrected flow labels of Liu et al.~\cite{liu2022error}. The CICDDoS2019 set uses the official CICFlowMeter v3 labels from the seven \texttt{03-11} testing-day files: \texttt{Portmap.csv} (58 positive alerts), and \texttt{NetBIOS.csv}, \texttt{LDAP.csv}, \texttt{MSSQL.csv}, \texttt{UDP.csv}, \texttt{UDPLag.csv}, and \texttt{Syn.csv} (57 each). The 1,600 benign alerts come from \texttt{BENIGN}-labeled flows in those same files. January 12 captures are excluded.

We filter nonfinite or malformed flow records, including negative durations and zero packet counts. To reduce repeated burst traffic, we collapse consecutive alerts with the same source IP, destination IP, destination port, protocol, and signature within a 60-second sliding window. We deterministically stratify positive alerts by attack subtype or phase. Benign alerts are sampled without replacement from the same capture days and stratified by background application protocol. Each set has a 1:4 malicious-to-benign ratio. Raw timestamps, row indices, and dataset-specific subnet identifiers are stripped from the initial alert payload. During full investigations, the helper agents can retrieve packet evidence beyond the initial alert payload (Section~\ref{subsec:framework}).

To associate alerts with flows in the CSE-CIC-IDS2018 DDoS evaluation, we match the direction-independent five-tuple and require the alert timestamp to be within the flow interval with a two-second tolerance. When multiple rows match, we select the row with the closest flow start. Labels are reserved for scoring. Signature cues remain available and can contribute to classification.

The alert-only control and full workflow receive the same initial evidence. The control returns a verdict in one Triage Agent step with follow-up questions and the Critique Agent disabled. In the full workflow, the Triage Agent can ask follow-up questions and the Critique Agent can request further investigation. If the full workflow ends with an unparseable verdict, we restart the investigation from the same alert and score the first parseable verdict from a completed run.

We compute F1, recall, precision, false-positive rate (FPR), and Matthews correlation coefficient (MCC) from each run's confusion matrix~\cite{chicco2020advantages}. We report means and sample standard deviations over three runs. Trained variants use independent training seeds 158, 487, and 763. Pretrained baselines repeat inference three times. Steps are mean Triage Agent steps per alert. Precision depends on the proportion of malicious alerts, which we set to 20\% in these evaluations. Operational alert streams can have a much smaller malicious fraction~\cite{alahmadi202299}.

\subsection{Performance and Transfer}
\label{subsec:eval-comparison}

\begin{figure}[!htbp]
\centering
\includegraphics[width=\textwidth]{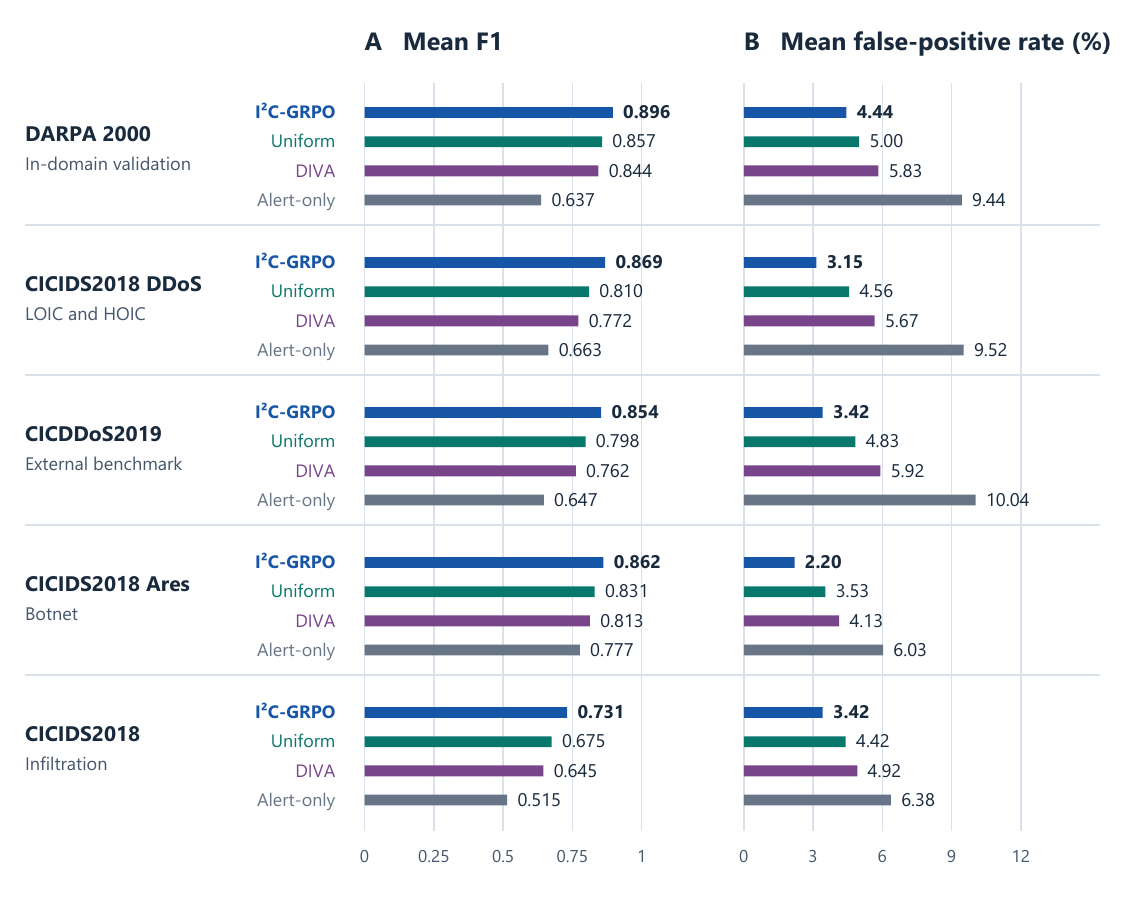}
\caption{Mean F1 and false-positive rate across three runs. Uniform weighting uses the same checklist and matcher as the proposed method. Alert-only triage uses the same initial evidence and returns a verdict in one step without Critique.}
\label{fig:benchmark-performance}
\end{figure}

The proposed method achieves the highest mean F1 on all five sets (Figure~\ref{fig:benchmark-performance}). Appendix~\ref{app:full-results} reports all configurations. On the CSE-CIC-IDS2018 DDoS set it reaches 0.869 F1 with a mean FPR of 3.15\%, compared with 0.701 F1 and 10.27\% FPR for GPT-5-mini in the full workflow. On CICDDoS2019 it reaches 0.854 F1, 5.7 percentage points above uniform weighting and 9.2 above DIVA.

The full workflow improves on alert-only triage in all five scenarios. On Ares, F1 rises from 0.777 to 0.862, with recall rising from 0.788 to 0.824. On infiltration, F1 rises from 0.515 to 0.731 and recall from 0.435 to 0.655. Infiltration shows the largest F1 gain among the four external tests. The control measures the combined effect of follow-up questions and Critique.

Mean paired F1 differences range from 3.1 to 5.8 points against uniform weighting and from 4.9 to 9.7 points against DIVA. Figure~\ref{fig:paired-f1-gains} reports two-sided 95\% paired $t$ intervals computed from the three seed-level differences ($df=2$). The intervals are unadjusted for multiple comparisons and describe variation across seeds on fixed alerts. The lower bound against uniform weighting on CSE-CIC-IDS2018 DDoS is approximately 0.09 percentage points, so this comparison is sensitive to training-seed variation. Variation across attack campaigns is outside the scope of these intervals.

\subsection{Reward Components, Checklist Order, and Insight Sources}
\label{subsec:reward-ablation}
\label{subsec:checklist-comparison}

Removing the insight reward produces the largest F1 reduction among the ablations on every scenario. On the CSE-CIC-IDS2018 DDoS set, F1 drops from 0.869 to 0.308 while the mean investigation length rises from 3.70 to 4.47 steps. Removing the question reward lowers F1 to 0.769 on this set.

Consensus ordering achieves higher F1 than randomized or reversed ordering under quadratic weighting on each scenario. In the randomized variant, a fresh checklist permutation is drawn for each prompt at each training step and shared by all candidates for that prompt. The reversed variant flips the consensus ordering. Uniform weighting also yields higher mean F1 than randomized quadratic weighting on all five sets. On infiltration, mean F1 is 0.731 with consensus quadratic weights, 0.675 with uniform weights, and 0.664 with randomized quadratic weights. Among single-expert variants, GPT-5.5 insights yield the highest mean F1 on all five sets.

Using GPT-5-mini as the sole source of checklist proposals also improves held-out triage. Qwen3-4B trained with these quadratically weighted checklists achieves mean F1 gains of 5.7 to 8.8 percentage points over the GPT-5-mini Triage Agent across the four external test sets. On infiltration, F1 rises from 0.594 to 0.664. GPT-5-mini generated the training trajectories in this ablation as well (Section~\ref{subsec:candidate}). Appendix~\ref{app:full-results} gives the full results.

\subsection{Sensitivity to the Importance Exponent}
\label{subsec:exponent}

Uniform, linear, and quadratic weighting form the $z=0,1,2$ family in Section~\ref{subsubsec:ranking}. With ten insights, their orness values are 0.50, 0.67, and 0.76 (Table~\ref{tab:orness}). Across the five evaluations, mean F1 increases with $z$ and mean FPR is lowest at $z=2$ (Table~\ref{tab:exponent}). Linear weighting increases FPR relative to uniform weighting on the two DDoS tests and DARPA evaluation, and reduces it on Ares and infiltration.

\begin{table}[t]
\centering
\small
\caption{Rank-exponent sensitivity. Cells give mean F1 and mean FPR (\%) across three seeds.}
\label{tab:exponent}
\begin{tabular}{@{}lccc@{}}
\toprule
Evaluation & $z=0$ & $z=1$ & $z=2$ \\
\midrule
DARPA 2000 & 0.857 / 5.00 & 0.865 / 6.39 & 0.896 / 4.44 \\
CICIDS2018 DDoS & 0.810 / 4.56 & 0.840 / 4.73 & 0.869 / 3.15 \\
CICDDoS2019 & 0.798 / 4.83 & 0.827 / 5.00 & 0.854 / 3.42 \\
CICIDS2018 Ares & 0.831 / 3.53 & 0.845 / 3.27 & 0.862 / 2.20 \\
CICIDS2018 infiltration & 0.675 / 4.42 & 0.703 / 4.08 & 0.731 / 3.42 \\
\bottomrule
\end{tabular}
\end{table}

\subsection{Repeatability of Matching and Ranking}
\label{subsec:stability}

A separate diagnostic evaluated the repeatability of matching fixed investigation responses to insight checklists. It sampled 100 nonempty original responses from 79 investigations, with 20 judgments per response and configuration. The response, checklist, and prompt were fixed across repetitions, while sampling seeds varied. The configurations were \nolinkurl{Qwen/Qwen3.6-35B-A3B-FP8}, Gemma 4 31B IT (BF16), Mistral Small 3.2 24B (BF16), and Nemotron 3 Nano 30B A3B (BF16).

For each response with all 20 judgments valid, we averaged pairwise Jaccard agreement between the sets of matched insights; two empty sets were assigned agreement 1. Mean agreement was 0.678 for Qwen, 0.918 for Gemma, 0.670 for Mistral, and 0.434 for Nemotron, based on 79, 99, 95, and 75 complete responses, respectively. On the common subset of 56 complete responses, the corresponding means were 0.673, 0.934, 0.675, and 0.424. Gemma also had the lowest mean within-response standard deviation of the resulting insight reward (Eq.~\eqref{eq:insight-reward} with $z=2$; 0.053, compared with 0.165, 0.146, and 0.295 for Qwen, Mistral, and Nemotron). Based on its repeatability and reward stability, we selected Gemma 4 31B IT as the training matcher. Across all configurations, 7,929 of 8,000 planned judgments were valid after retries. The diagnostic compares repeatability on saved responses. Model identity, precision, and prompt templates vary across the configurations.

Gemma also had the highest mean Jaccard agreement after excluding judgment pairs for which both matched sets were empty (Figure~\ref{fig:matcher-repeatability}).

A separate ranking diagnostic sampled 100 training checklists and requested 20 rankings per checklist from DeepSeek V4 Pro with independently shuffled input order. One single-item checklist was excluded from the agreement analysis. Mean Kendall's $W$ across the remaining 99 checklists was 0.692 (Figure~\ref{fig:ranking-repeatability}). The estimate combines sampling and presentation-order effects for one ranker. It describes whole-list repeatability. Top-rank stability and agreement among the three panel rankers were not measured.

\FloatBarrier

\section{Discussion}
\label{sec:discussion}

The uniform and quadratic variants share a checklist and matcher, so their comparison tests the effect of changing rank weights. Quadratic weighting improves mean F1 even on Ares and infiltration, where the two variants use the same mean number of steps. Randomizing or reversing the ordering also reduces mean F1, suggesting that which insights receive the largest weights matters.

The GPT-5-mini ablation shows that the policy can improve on the trajectory generator with proposals from that same model. The extractor, rankers, and matcher still shape the supervision. The full panel achieves higher mean F1 than any single-expert variant, but the comparison also changes checklist content and potentially its length, so it cannot isolate the contribution of panel diversity.

Removing the question reward shortens investigations, lowers precision, and raises the false-positive rate on all five sets. Recall rises slightly on both external DDoS sets. On CSE-CIC-IDS2018 DDoS, recall rises from 86.3\% to 87.5\% and the false-positive rate rises from 3.15\% to 10.08\% (Appendix~\ref{app:full-results}).

The alert-only control shows the combined contribution of follow-up questions and verdict review. The full workflow reduces false positives on Ares and improves recall on infiltration, where it still misses about one-third of malicious alerts. A trace audit could establish which retrieved evidence changed individual verdicts.

\subsection{Limitations and Threats to Validity}
\label{subsec:limitations}

\paragraph{Reward validity and judge uncertainty.}
Coverage rewards the checklist entries that candidates target or assert. Unsupported claims can receive credit, and the reward has no penalty for extraneous findings. We did not track how many findings the matcher extracted or how reasoning length changed during training. Group standardization can amplify small score differences (Eq.~\eqref{eq:grpo-advantage}), and length normalization can affect the policy update~\cite{liu2025understanding}. The diagnostics measure repeatability on fixed inputs. They leave judge correctness and agreement among panel rankers unmeasured. Binary matching gives no partial credit, and random tie-breaking can assign unequal weights to equally ranked insights.

\paragraph{Scope and possible shortcuts.}
The DARPA result pools held-out alerts from the training campaign with alerts from a second campaign. The external sets test transfer to specific laboratory captures at a fixed class prevalence. Visible signatures may provide classification shortcuts, and pretrained models may recognize publicly documented datasets. The paired intervals quantify variation across training seeds on the same alerts. Generalization across campaigns and networks requires independent captures.

\paragraph{Evaluation and reproducibility.}
The results summarize parseable verdicts. Full investigation restarts may add work, and restart and Critique rejection rates were unavailable by variant. Deployment uses the fixed helper models. The offline expert and ranker panels supply supervision during training. Releasing the checklists and contexts would allow the reward data to be reused even if hosted model revisions change.

\section{Conclusion}
\label{sec:conclusion}

I$^2$C-GRPO trains a Triage Agent on recorded investigation prefixes by scoring candidate actions against ranked forensic insights without executing them. Quadratic weighting yields higher mean F1 than uniform weighting and DIVA on all five evaluation sets. The GPT-5-mini ablation shows that gains on external sets are possible when the trajectory generator also supplies the checklist proposals.

The full investigation improves F1 over one-step triage on all five sets, with the largest external gain on infiltration. Recall on that set still trails the other external sets. Future evaluation should audit which retrieved evidence changes verdicts and test whether graded matching and ranker disagreement improve the training signal across independent campaigns.

\appendix
\section{Full Evaluation Results}
\label{app:full-results}
The first three rows of each table are pretrained baselines. All trained variants use Qwen3-4B as the Triage Agent base. The positive class consists of malicious alerts.
\begin{figure}[!htbp]
\centering
\includegraphics[width=\textwidth]{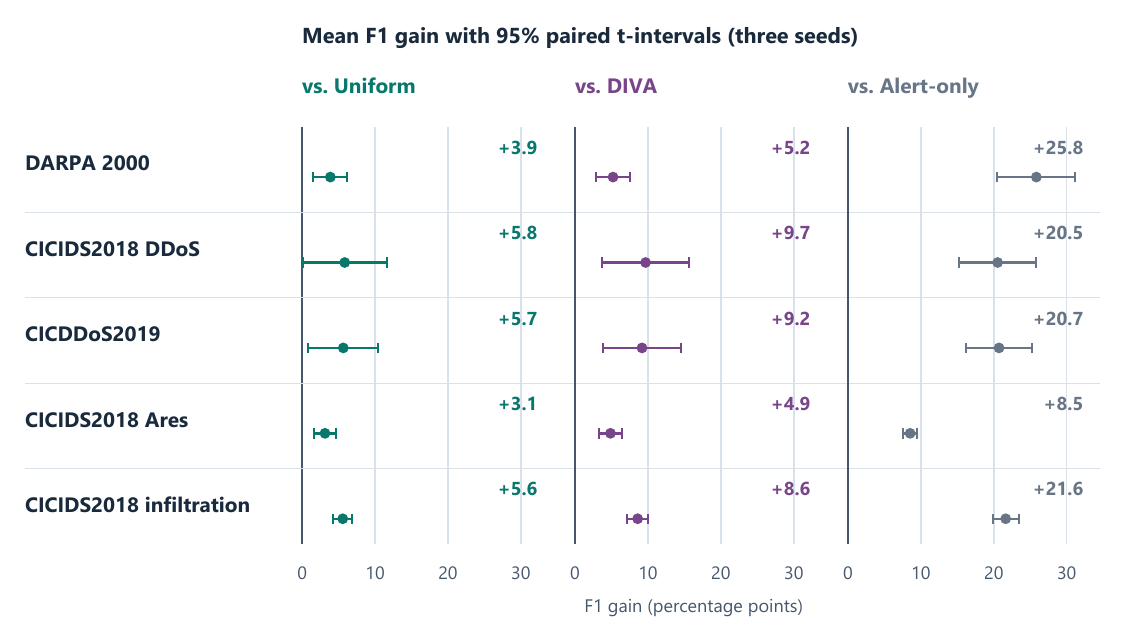}
\caption{I$^2$C-GRPO F1 gains over uniform weighting, DIVA, and alert-only triage. Marks show mean differences in percentage points and bars show two-sided 95\% paired $t$ intervals across the three seed-level differences.}
\label{fig:paired-f1-gains}
\end{figure}

\begin{table*}[t]
\centering
\scriptsize
\setlength{\tabcolsep}{4pt}
\caption{DARPA 2000. Values are mean (sample SD) across three runs. FPR is in percent.}
\label{tab:comparison}
\resizebox{\textwidth}{!}{%
\begin{tabular}{@{}lcccccc@{}}
\toprule
Configuration & F1 & Recall & Precision & FPR & MCC & Steps \\
\midrule
Qwen3-4B & $0.244(0.001)$ & $0.478(0.019)$ & $0.164(0.001)$ & $60.83(2.89)$ & $-0.106(0.009)$ & $1.83(0.06)$ \\
Qwen3.5-35B-A3B & $0.542(0.066)$ & $0.633(0.058)$ & $0.473(0.069)$ & $17.78(3.15)$ & $0.412(0.088)$ & $2.73(0.15)$ \\
GPT-5-mini & $0.720(0.039)$ & $0.789(0.164)$ & $0.714(0.207)$ & $10.28(8.55)$ & $0.666(0.061)$ & $2.97(0.32)$ \\
SFT & $0.544(0.104)$ & $0.478(0.096)$ & $0.630(0.113)$ & $6.94(1.92)$ & $0.455(0.125)$ & $2.63(0.29)$ \\
Alert-only & $0.637(0.013)$ & $0.644(0.019)$ & $0.631(0.022)$ & $9.44(0.96)$ & $0.546(0.016)$ & $1.00(0.00)$ \\
dDPO & $0.685(0.013)$ & $0.711(0.019)$ & $0.662(0.036)$ & $9.17(1.67)$ & $0.603(0.019)$ & $3.00(0.10)$ \\
RLCF & $0.710(0.018)$ & $0.733(0.033)$ & $0.688(0.018)$ & $8.33(0.83)$ & $0.635(0.022)$ & $3.07(0.15)$ \\
dRLAIF & $0.762(0.034)$ & $0.811(0.019)$ & $0.720(0.069)$ & $8.06(2.68)$ & $0.700(0.044)$ & $3.20(0.10)$ \\
DIVA & $0.844(0.005)$ & $0.900(0.033)$ & $0.795(0.017)$ & $5.83(0.83)$ & $0.804(0.007)$ & $3.23(0.15)$ \\
Uniform & $0.857(0.005)$ & $0.900(0.033)$ & $0.819(0.019)$ & $5.00(0.83)$ & $0.821(0.006)$ & $3.30(0.10)$ \\
Linear & $0.865(0.021)$ & $0.956(0.019)$ & $0.792(0.049)$ & $6.39(1.92)$ & $0.834(0.024)$ & $3.33(0.12)$ \\
GPT-5.5 insights only & $0.881(0.007)$ & $0.944(0.019)$ & $0.826(0.022)$ & $5.00(0.83)$ & $0.852(0.009)$ & $3.37(0.06)$ \\
GPT-5-mini insights only & $0.812(0.006)$ & $0.867(0.033)$ & $0.765(0.016)$ & $6.67(0.83)$ & $0.765(0.008)$ & $3.37(0.06)$ \\
Claude-only insights & $0.871(0.005)$ & $0.933(0.033)$ & $0.817(0.031)$ & $5.28(1.27)$ & $0.839(0.004)$ & $3.47(0.06)$ \\
Gemini-only insights & $0.860(0.003)$ & $0.922(0.019)$ & $0.806(0.010)$ & $5.56(0.48)$ & $0.825(0.004)$ & $3.57(0.06)$ \\
Randomized order & $0.848(0.004)$ & $0.900(0.033)$ & $0.804(0.032)$ & $5.56(1.27)$ & $0.810(0.006)$ & $3.63(0.15)$ \\
Reversed order & $0.825(0.006)$ & $0.867(0.033)$ & $0.789(0.018)$ & $5.83(0.83)$ & $0.781(0.007)$ & $3.87(0.15)$ \\
No insight reward & $0.310(0.007)$ & $0.433(0.033)$ & $0.241(0.011)$ & $34.17(4.41)$ & $0.077(0.015)$ & $4.07(0.15)$ \\
No question reward & $0.789(0.006)$ & $0.933(0.033)$ & $0.683(0.009)$ & $10.83(0.83)$ & $0.741(0.011)$ & $2.80(0.10)$ \\
\ac{I2C}-GRPO & $0.896(0.009)$ & $0.956(0.019)$ & $0.843(0.013)$ & $4.44(0.48)$ & $0.871(0.012)$ & $3.27(0.06)$ \\
\bottomrule
\end{tabular}}
\end{table*}

\begin{table*}[t]
\centering
\scriptsize
\setlength{\tabcolsep}{4pt}
\caption{CICIDS2018 DDoS. Values are mean (sample SD) across three runs. FPR is in percent.}
\label{tab:results-loic}
\resizebox{\textwidth}{!}{%
\begin{tabular}{@{}lcccccc@{}}
\toprule
Configuration & F1 & Recall & Precision & FPR & MCC & Steps \\
\midrule
Qwen3-4B & $0.236(0.009)$ & $0.455(0.003)$ & $0.159(0.009)$ & $60.10(4.15)$ & $-0.118(0.033)$ & $1.97(0.06)$ \\
Qwen3.5-35B-A3B & $0.522(0.014)$ & $0.620(0.005)$ & $0.450(0.024)$ & $18.98(1.89)$ & $0.385(0.020)$ & $2.90(0.10)$ \\
GPT-5-mini & $0.701(0.017)$ & $0.760(0.008)$ & $0.650(0.035)$ & $10.27(1.59)$ & $0.621(0.022)$ & $3.20(0.10)$ \\
SFT & $0.609(0.002)$ & $0.580(0.010)$ & $0.640(0.010)$ & $8.15(0.48)$ & $0.518(0.002)$ & $2.90(0.10)$ \\
Alert-only & $0.663(0.002)$ & $0.685(0.010)$ & $0.643(0.007)$ & $9.52(0.42)$ & $0.576(0.002)$ & $1.00(0.00)$ \\
dDPO & $0.649(0.008)$ & $0.680(0.013)$ & $0.620(0.024)$ & $10.44(1.20)$ & $0.557(0.012)$ & $3.20(0.10)$ \\
RLCF & $0.679(0.008)$ & $0.705(0.015)$ & $0.654(0.021)$ & $9.33(1.00)$ & $0.595(0.011)$ & $3.30(0.10)$ \\
dRLAIF & $0.734(0.023)$ & $0.785(0.018)$ & $0.690(0.054)$ & $8.94(2.27)$ & $0.664(0.031)$ & $3.50(0.10)$ \\
DIVA & $0.772(0.008)$ & $0.771(0.019)$ & $0.773(0.007)$ & $5.67(0.32)$ & $0.715(0.009)$ & $3.60(0.10)$ \\
Uniform & $0.810(0.007)$ & $0.805(0.015)$ & $0.815(0.004)$ & $4.56(0.17)$ & $0.763(0.008)$ & $3.77(0.06)$ \\
Linear & $0.840(0.014)$ & $0.860(0.020)$ & $0.821(0.041)$ & $4.73(1.34)$ & $0.799(0.019)$ & $3.80(0.10)$ \\
GPT-5.5 insights only & $0.855(0.013)$ & $0.847(0.023)$ & $0.865(0.027)$ & $3.33(0.79)$ & $0.820(0.016)$ & $3.80(0.10)$ \\
GPT-5-mini insights only & $0.789(0.008)$ & $0.782(0.018)$ & $0.797(0.008)$ & $5.00(0.33)$ & $0.737(0.009)$ & $3.70(0.10)$ \\
Claude-only insights & $0.838(0.013)$ & $0.827(0.023)$ & $0.851(0.026)$ & $3.65(0.79)$ & $0.799(0.016)$ & $3.90(0.10)$ \\
Gemini-only insights & $0.825(0.013)$ & $0.812(0.025)$ & $0.838(0.025)$ & $3.96(0.79)$ & $0.782(0.016)$ & $4.00(0.10)$ \\
Randomized order & $0.795(0.006)$ & $0.787(0.013)$ & $0.802(0.004)$ & $4.85(0.16)$ & $0.744(0.007)$ & $4.03(0.15)$ \\
Reversed order & $0.749(0.007)$ & $0.738(0.013)$ & $0.762(0.005)$ & $5.77(0.16)$ & $0.688(0.009)$ & $4.27(0.15)$ \\
No insight reward & $0.308(0.012)$ & $0.500(0.022)$ & $0.222(0.015)$ & $43.92(5.06)$ & $0.049(0.029)$ & $4.47(0.15)$ \\
No question reward & $0.769(0.018)$ & $0.875(0.025)$ & $0.686(0.039)$ & $10.08(1.97)$ & $0.711(0.021)$ & $3.10(0.10)$ \\
\ac{I2C}-GRPO & $0.869(0.022)$ & $0.863(0.029)$ & $0.876(0.061)$ & $3.15(1.71)$ & $0.837(0.030)$ & $3.70(0.10)$ \\
\bottomrule
\end{tabular}}
\end{table*}

\begin{table*}[t]
\centering
\scriptsize
\setlength{\tabcolsep}{4pt}
\caption{CICDDoS2019. Values are mean (sample SD) across three runs. FPR is in percent.}
\label{tab:results-cicddos}
\resizebox{\textwidth}{!}{%
\begin{tabular}{@{}lcccccc@{}}
\toprule
Configuration & F1 & Recall & Precision & FPR & MCC & Steps \\
\midrule
Qwen3-4B & $0.241(0.006)$ & $0.450(0.005)$ & $0.164(0.006)$ & $57.29(3.08)$ & $-0.099(0.022)$ & $1.93(0.06)$ \\
Qwen3.5-35B-A3B & $0.537(0.013)$ & $0.635(0.005)$ & $0.466(0.022)$ & $18.23(1.72)$ & $0.407(0.019)$ & $2.83(0.15)$ \\
GPT-5-mini & $0.715(0.013)$ & $0.771(0.009)$ & $0.667(0.029)$ & $9.65(1.32)$ & $0.640(0.018)$ & $3.00(0.26)$ \\
SFT & $0.619(0.002)$ & $0.589(0.010)$ & $0.653(0.009)$ & $7.83(0.44)$ & $0.531(0.002)$ & $2.70(0.26)$ \\
Alert-only & $0.647(0.001)$ & $0.670(0.010)$ & $0.625(0.008)$ & $10.04(0.51)$ & $0.555(0.002)$ & $1.00(0.00)$ \\
dDPO & $0.642(0.005)$ & $0.670(0.015)$ & $0.617(0.019)$ & $10.42(1.02)$ & $0.549(0.008)$ & $3.03(0.12)$ \\
RLCF & $0.669(0.006)$ & $0.695(0.015)$ & $0.646(0.018)$ & $9.54(0.90)$ & $0.584(0.009)$ & $3.10(0.10)$ \\
dRLAIF & $0.723(0.020)$ & $0.772(0.018)$ & $0.682(0.048)$ & $9.12(2.06)$ & $0.651(0.027)$ & $3.27(0.15)$ \\
DIVA & $0.762(0.010)$ & $0.762(0.020)$ & $0.763(0.005)$ & $5.92(0.26)$ & $0.703(0.011)$ & $3.30(0.10)$ \\
Uniform & $0.798(0.009)$ & $0.792(0.018)$ & $0.804(0.005)$ & $4.83(0.19)$ & $0.748(0.011)$ & $3.37(0.06)$ \\
Linear & $0.827(0.012)$ & $0.845(0.020)$ & $0.810(0.035)$ & $5.00(1.19)$ & $0.783(0.016)$ & $3.40(0.10)$ \\
GPT-5.5 insights only & $0.841(0.011)$ & $0.832(0.023)$ & $0.852(0.023)$ & $3.62(0.70)$ & $0.803(0.014)$ & $3.40(0.10)$ \\
GPT-5-mini insights only & $0.779(0.008)$ & $0.772(0.018)$ & $0.786(0.008)$ & $5.25(0.33)$ & $0.725(0.009)$ & $3.30(0.10)$ \\
Claude-only insights & $0.823(0.011)$ & $0.812(0.023)$ & $0.836(0.022)$ & $4.00(0.70)$ & $0.780(0.014)$ & $3.50(0.10)$ \\
Gemini-only insights & $0.809(0.012)$ & $0.795(0.025)$ & $0.824(0.022)$ & $4.25(0.70)$ & $0.763(0.014)$ & $3.60(0.10)$ \\
Randomized order & $0.783(0.008)$ & $0.775(0.015)$ & $0.791(0.004)$ & $5.12(0.12)$ & $0.729(0.010)$ & $3.70(0.10)$ \\
Reversed order & $0.736(0.009)$ & $0.725(0.015)$ & $0.747(0.004)$ & $6.12(0.12)$ & $0.671(0.010)$ & $3.97(0.15)$ \\
No insight reward & $0.307(0.012)$ & $0.492(0.023)$ & $0.223(0.015)$ & $42.92(4.73)$ & $0.051(0.028)$ & $4.13(0.12)$ \\
No question reward & $0.757(0.015)$ & $0.860(0.025)$ & $0.677(0.035)$ & $10.33(1.81)$ & $0.696(0.018)$ & $2.87(0.06)$ \\
\ac{I2C}-GRPO & $0.854(0.019)$ & $0.847(0.028)$ & $0.864(0.052)$ & $3.42(1.49)$ & $0.819(0.026)$ & $3.33(0.06)$ \\
\bottomrule
\end{tabular}}
\end{table*}

\begin{table*}[t]
\centering
\scriptsize
\setlength{\tabcolsep}{4pt}
\caption{CICIDS2018 Ares. Values are mean (sample SD) across three runs. FPR is in percent.}
\label{tab:results-ares}
\resizebox{\textwidth}{!}{%
\begin{tabular}{@{}lcccccc@{}}
\toprule
Configuration & F1 & Recall & Precision & FPR & MCC & Steps \\
\midrule
Qwen3-4B & $0.310(0.005)$ & $0.511(0.010)$ & $0.223(0.006)$ & $44.50(2.18)$ & $0.053(0.011)$ & $1.77(0.06)$ \\
Qwen3.5-35B-A3B & $0.575(0.003)$ & $0.656(0.016)$ & $0.512(0.011)$ & $15.67(1.04)$ & $0.458(0.004)$ & $2.57(0.15)$ \\
GPT-5-mini & $0.752(0.008)$ & $0.784(0.016)$ & $0.722(0.022)$ & $7.57(0.93)$ & $0.687(0.011)$ & $2.73(0.21)$ \\
SFT & $0.689(0.006)$ & $0.679(0.018)$ & $0.700(0.007)$ & $7.27(0.42)$ & $0.613(0.006)$ & $2.43(0.21)$ \\
Alert-only & $0.777(0.003)$ & $0.788(0.012)$ & $0.766(0.009)$ & $6.03(0.38)$ & $0.720(0.003)$ & $1.00(0.00)$ \\
dDPO & $0.735(0.004)$ & $0.744(0.016)$ & $0.727(0.011)$ & $7.00(0.53)$ & $0.668(0.004)$ & $2.80(0.10)$ \\
RLCF & $0.760(0.006)$ & $0.768(0.016)$ & $0.753(0.010)$ & $6.30(0.44)$ & $0.700(0.007)$ & $2.90(0.10)$ \\
dRLAIF & $0.790(0.007)$ & $0.800(0.016)$ & $0.780(0.021)$ & $5.67(0.76)$ & $0.736(0.010)$ & $3.00(0.10)$ \\
DIVA & $0.813(0.008)$ & $0.799(0.018)$ & $0.829(0.009)$ & $4.13(0.31)$ & $0.768(0.009)$ & $3.03(0.12)$ \\
Uniform & $0.831(0.008)$ & $0.811(0.018)$ & $0.852(0.009)$ & $3.53(0.31)$ & $0.790(0.009)$ & $3.10(0.10)$ \\
Linear & $0.845(0.007)$ & $0.827(0.018)$ & $0.864(0.013)$ & $3.27(0.42)$ & $0.807(0.008)$ & $3.17(0.06)$ \\
GPT-5.5 insights only & $0.848(0.008)$ & $0.820(0.020)$ & $0.878(0.014)$ & $2.87(0.42)$ & $0.812(0.010)$ & $3.20(0.10)$ \\
GPT-5-mini insights only & $0.809(0.008)$ & $0.788(0.020)$ & $0.831(0.009)$ & $4.00(0.35)$ & $0.763(0.008)$ & $3.10(0.10)$ \\
Claude-only insights & $0.835(0.008)$ & $0.808(0.020)$ & $0.865(0.014)$ & $3.17(0.42)$ & $0.797(0.010)$ & $3.30(0.10)$ \\
Gemini-only insights & $0.823(0.009)$ & $0.796(0.020)$ & $0.852(0.013)$ & $3.47(0.42)$ & $0.781(0.010)$ & $3.40(0.10)$ \\
Randomized order & $0.817(0.008)$ & $0.799(0.018)$ & $0.836(0.009)$ & $3.93(0.31)$ & $0.772(0.009)$ & $3.47(0.15)$ \\
Reversed order & $0.786(0.008)$ & $0.771(0.018)$ & $0.803(0.008)$ & $4.73(0.31)$ & $0.735(0.010)$ & $3.67(0.15)$ \\
No insight reward & $0.387(0.008)$ & $0.560(0.020)$ & $0.296(0.009)$ & $33.33(2.08)$ & $0.187(0.013)$ & $3.87(0.15)$ \\
No question reward & $0.798(0.006)$ & $0.824(0.016)$ & $0.775(0.013)$ & $6.00(0.53)$ & $0.747(0.008)$ & $2.60(0.10)$ \\
\ac{I2C}-GRPO & $0.862(0.007)$ & $0.824(0.016)$ & $0.904(0.020)$ & $2.20(0.53)$ & $0.831(0.009)$ & $3.10(0.10)$ \\
\bottomrule
\end{tabular}}
\end{table*}

\begin{table*}[t]
\centering
\scriptsize
\setlength{\tabcolsep}{4pt}
\caption{CICIDS2018 infiltration. Values are mean (sample SD) across three runs. FPR is in percent.}
\label{tab:results-infiltration}
\resizebox{\textwidth}{!}{%
\begin{tabular}{@{}lcccccc@{}}
\toprule
Configuration & F1 & Recall & Precision & FPR & MCC & Steps \\
\midrule
Qwen3-4B & $0.175(0.004)$ & $0.280(0.010)$ & $0.128(0.004)$ & $47.92(2.60)$ & $-0.161(0.015)$ & $2.20(0.10)$ \\
Qwen3.5-35B-A3B & $0.415(0.007)$ & $0.425(0.015)$ & $0.406(0.010)$ & $15.54(1.02)$ & $0.265(0.009)$ & $3.17(0.15)$ \\
GPT-5-mini & $0.594(0.008)$ & $0.550(0.020)$ & $0.647(0.022)$ & $7.54(0.90)$ & $0.505(0.010)$ & $3.43(0.21)$ \\
SFT & $0.478(0.012)$ & $0.410(0.020)$ & $0.575(0.006)$ & $7.58(0.52)$ & $0.382(0.009)$ & $3.07(0.15)$ \\
Alert-only & $0.515(0.009)$ & $0.435(0.015)$ & $0.631(0.005)$ & $6.38(0.33)$ & $0.431(0.007)$ & $1.00(0.00)$ \\
dDPO & $0.539(0.009)$ & $0.480(0.020)$ & $0.616(0.012)$ & $7.50(0.66)$ & $0.447(0.006)$ & $3.40(0.10)$ \\
RLCF & $0.572(0.012)$ & $0.510(0.020)$ & $0.651(0.013)$ & $6.83(0.52)$ & $0.486(0.012)$ & $3.50(0.10)$ \\
dRLAIF & $0.622(0.010)$ & $0.560(0.020)$ & $0.701(0.019)$ & $6.00(0.66)$ & $0.546(0.011)$ & $3.70(0.10)$ \\
DIVA & $0.645(0.012)$ & $0.570(0.020)$ & $0.744(0.012)$ & $4.92(0.38)$ & $0.578(0.012)$ & $3.77(0.15)$ \\
Uniform & $0.675(0.012)$ & $0.600(0.020)$ & $0.773(0.012)$ & $4.42(0.38)$ & $0.614(0.012)$ & $3.90(0.10)$ \\
Linear & $0.703(0.010)$ & $0.630(0.020)$ & $0.795(0.018)$ & $4.08(0.52)$ & $0.645(0.011)$ & $3.97(0.15)$ \\
GPT-5.5 insights only & $0.718(0.010)$ & $0.640(0.020)$ & $0.818(0.019)$ & $3.58(0.52)$ & $0.665(0.011)$ & $4.00(0.10)$ \\
GPT-5-mini insights only & $0.664(0.012)$ & $0.590(0.020)$ & $0.760(0.012)$ & $4.67(0.38)$ & $0.600(0.012)$ & $3.87(0.15)$ \\
Claude-only insights & $0.699(0.010)$ & $0.620(0.020)$ & $0.802(0.019)$ & $3.83(0.52)$ & $0.644(0.011)$ & $4.10(0.10)$ \\
Gemini-only insights & $0.680(0.011)$ & $0.600(0.020)$ & $0.787(0.018)$ & $4.08(0.52)$ & $0.622(0.011)$ & $4.20(0.10)$ \\
Randomized order & $0.664(0.012)$ & $0.590(0.020)$ & $0.760(0.012)$ & $4.67(0.38)$ & $0.600(0.012)$ & $4.30(0.10)$ \\
Reversed order & $0.615(0.013)$ & $0.540(0.020)$ & $0.714(0.011)$ & $5.42(0.38)$ & $0.542(0.012)$ & $4.50(0.10)$ \\
No insight reward & $0.244(0.008)$ & $0.350(0.020)$ & $0.188(0.007)$ & $37.92(2.60)$ & $-0.024(0.015)$ & $4.67(0.15)$ \\
No question reward & $0.610(0.010)$ & $0.580(0.020)$ & $0.645(0.014)$ & $8.00(0.66)$ & $0.521(0.011)$ & $3.20(0.10)$ \\
\ac{I2C}-GRPO & $0.731(0.011)$ & $0.655(0.025)$ & $0.829(0.031)$ & $3.42(0.80)$ & $0.681(0.013)$ & $3.90(0.10)$ \\
\bottomrule
\end{tabular}}
\end{table*}

\FloatBarrier
\subsection{Matching and Ranking Repeatability}

\begin{figure}[!htbp]
\centering
\includegraphics[width=\textwidth]{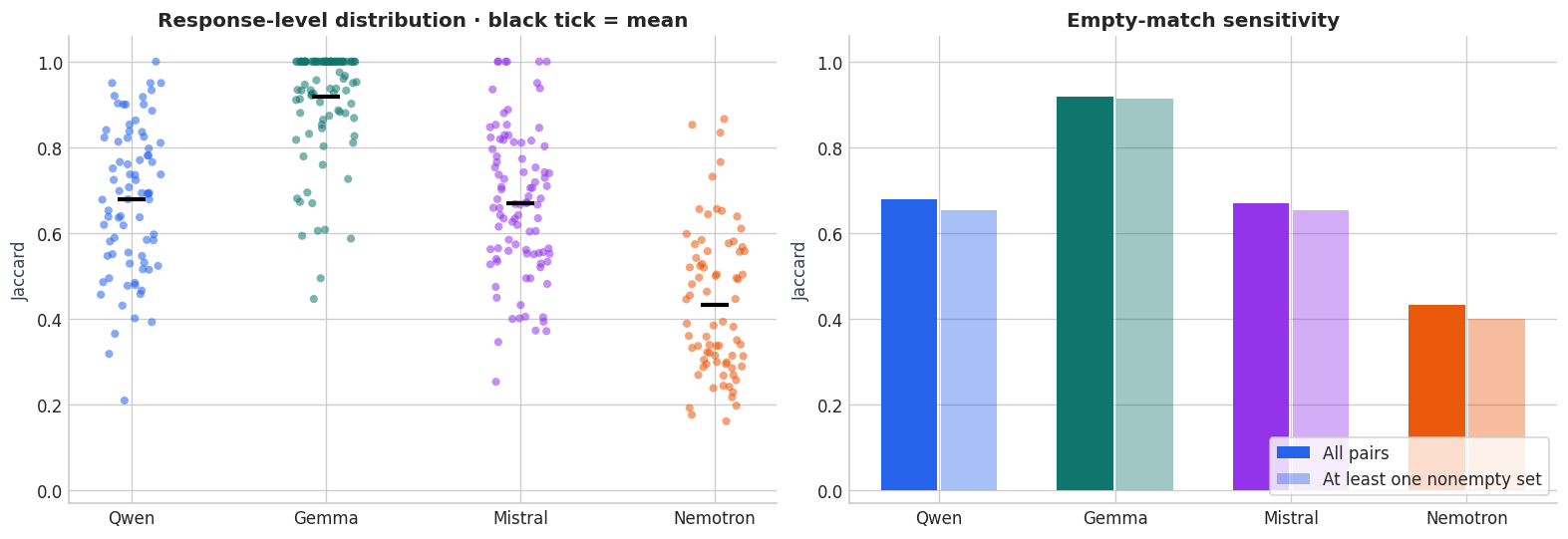}
\caption{Repeatability of insight matching on fixed responses. In the left panel, each dot is the mean pairwise Jaccard agreement for one response with 20 valid judgments, and black ticks mark means across responses. The right panel compares agreement using all judgment pairs with agreement after excluding pairs in which both matched sets are empty.}
\label{fig:matcher-repeatability}
\end{figure}

\begin{figure}[!htbp]
\centering
\includegraphics[width=0.94\textwidth]{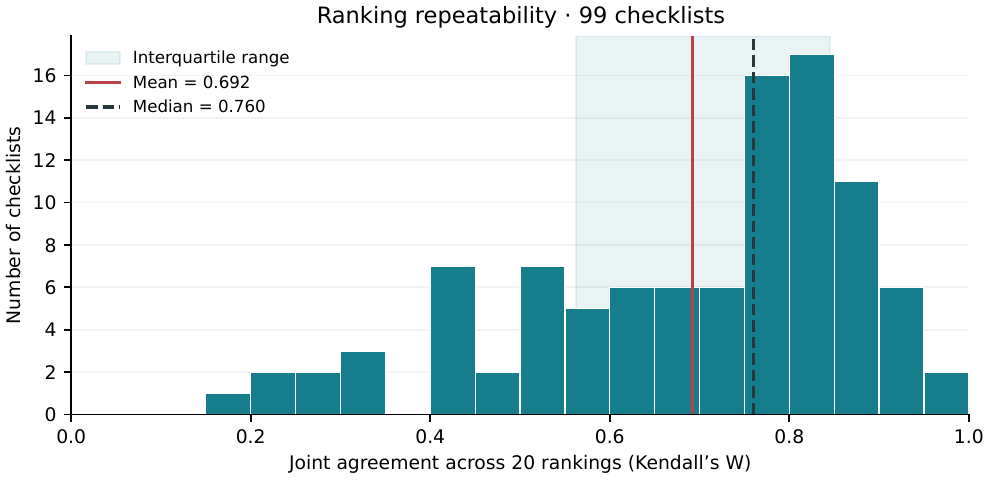}
\caption{Distribution of Kendall's $W$ across 99 checklists, each ranked 20 times by DeepSeek V4 Pro with independently shuffled input order. Solid and dashed lines mark the mean (0.692) and median (0.760), respectively. The shaded band spans the interquartile range (0.562--0.845).}
\label{fig:ranking-repeatability}
\end{figure}

\FloatBarrier

\section*{CRediT authorship contribution statement}
\textbf{Eyal Zloof:} Conceptualization, Methodology, Software, Investigation, Writing -- original draft, Writing -- review \& editing.
\textbf{Shaun Shuster:} Conceptualization, Methodology, Software, Investigation, Writing -- original draft, Writing -- review \& editing.
\textbf{Aviram Zilberman:} Writing -- review \& editing.
\textbf{Chen Frydman:} Data curation, Writing -- review \& editing.
\textbf{Inderjeet Singh:} Writing -- review \& editing.
\textbf{Andr\'{e}s Murillo:} Writing -- review \& editing.
\textbf{Yosi Hadad:} Data curation, Writing -- review \& editing.
\textbf{Asaf Shabtai:} Writing -- review \& editing -- Supervision.
\textbf{Yuval Elovici:} Writing -- review \& editing -- Supervision.
\textbf{Rami Puzis:} Writing -- review \& editing -- Supervision.

\section*{Declaration of competing interest}
The authors declare that they have no known competing financial interests or personal relationships that could have appeared to influence the work reported in this paper.

\section*{Data availability}
The DARPA 2000 LLDOS 1.0 and 2.0.2 datasets are publicly available from MIT Lincoln Laboratory at \url{https://www.ll.mit.edu/r-d/datasets/2000-darpa-intrusion-detection-scenario-specific-datasets}, and the CSE-CIC-IDS2018 and CICDDoS2019 datasets from the Canadian Institute for Cybersecurity at \url{https://www.unb.ca/cic/datasets/ids-2018.html} and \url{https://www.unb.ca/cic/datasets/ddos-2019.html}, respectively. The code, curated investigation contexts, and ranked forensic-insight dataset produced in this work will be made publicly available upon publication.

\section*{Declaration of generative AI and AI-assisted technologies in the manuscript preparation process}
During the preparation of this work, the authors used Grammarly and Claude for language editing and OpenAI Codex to revise the manuscript and prepare figures from the reported results. After using these tools, the authors reviewed and edited the content as needed and take full responsibility for the content of the published article. The large language models used as components of the proposed method and baselines are described in Sections~\ref{sec:method} and~\ref{sec:evaluation}.

\bibliographystyle{elsarticle-num}
\bibliography{Bibliography/Bibliography}

\end{document}


\section{Threat Model}\label{sec:threat-model}

We consider \ac{DDoS} campaigns where an external adversary progresses from reconnaissance and vulnerability probing through host compromise, malware installation, and coordinated \ac{DDoS} execution, in which an external adversary progresses from reconnaissance and vulnerability probing to host compromise, malware installation, and coordinated \ac{DDoS} execution, thereby threatening network availability and internal compute infrastructure. 
We assume that a network intrusion detection system monitors the traffic using a standard set of rules. 
We further assume that \acp{PCAP} of recent traffic are available for inspection in unaltered form before being discarded.  
The purpose of alert triaging is to perform initial investigation of the alerts based on available traffic on behalf of the analyst and provide judgment whether the alert is likely an intrusion or a false positive.


\section{Overview of the Alert Triaging Workflow}
\label{subsec:framework}

The workflow consists of four agents that collaborate during the investigation process:

\begin{itemize}
    \item \textbf{Context Gathering Agent:} Aggregates and organizes the initial evidence from PCAP analysis, alert history, and inventory/IP mappings before the investigation begins.

    \item \textbf{Triage Agent:} The \emph{only} trainable component. Given the current investigation context, it generates investigative questions to guide the investigation or concludes it with a structured decision, including alert classification, confidence, and impact scores.

    \item \textbf{Question Answering Agent:} A fixed agent that answers query actions from the Triage Agent by querying the indexed packet-level evidence database built from the raw PCAP traces and a complementary \ac{KG}, returning the result as part of the updated context; it does not make decisions or coordinate other agents.

    \item \textbf{Critique Agent:} A fixed agent that assesses whether the Triage Agent's final classification, confidence, and reasoning are sufficiently justified by the accumulated evidence, rejecting unsupported or incomplete conclusions and requiring further investigation.

\end{itemize}

\begin{figure}
    \centering
    \includegraphics[width=\linewidth]{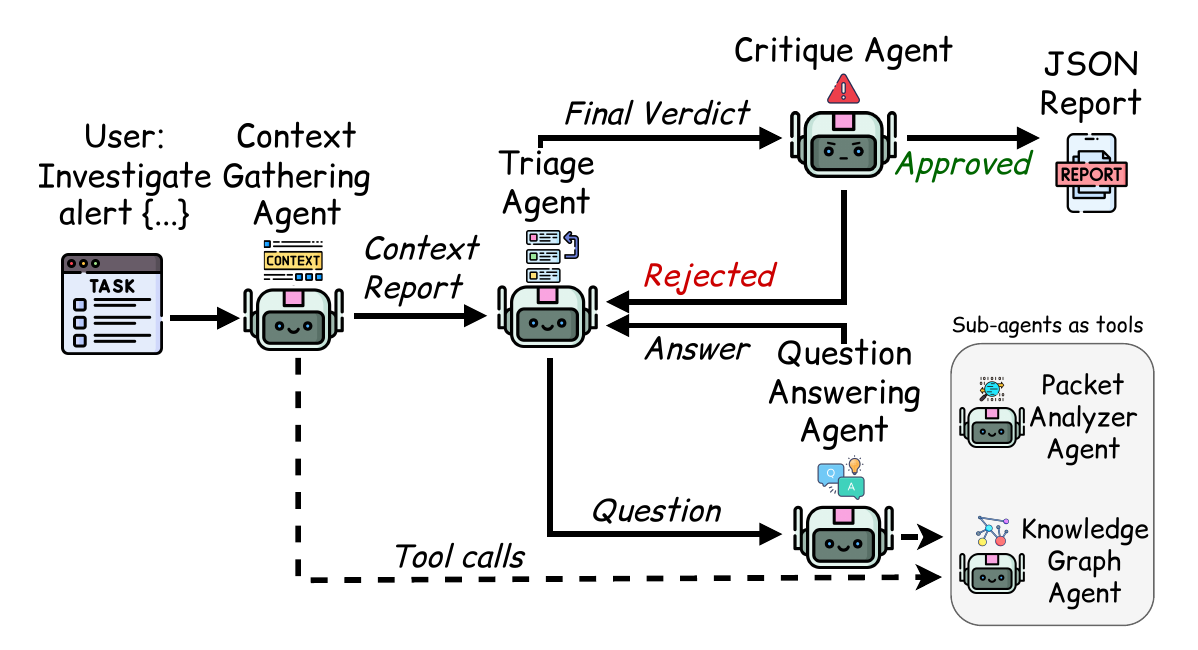}
    \caption{Thealerts triaging workflow}
    \label{fig:agent-workflow}
\end{figure}

Figure~\ref{fig:agent-workflow} shows the alert triaging workflow.

Only the Triage Agent is optimized during training; the remaining agents are fixed helper components implemented through static prompts and tool interfaces.

\paragraph{Knowledge Graph Evidence.}
The \ac{KG} is a read-only Neo4j evidence store used by the Context Gathering and Question Answering Agents for structured factual context, representing alerts, IP addresses, detection rules, and network segments as nodes linked by alert-endpoint, alert-rule, and host-segment relations. Agents retrieve host inventory, network topology, rule, and alert-history facts via constrained Cypher queries; packet- and stream-level questions are instead answered from the indexed raw-PCAP database. Graph queries cannot write data or access evaluation labels, verdicts, or benchmark-role metadata.

\begin{acronym}
  \acro{AI}{Artificial Intelligence}
  \acro{API}{Application Programming Interface}
  \acro{APT}{Advanced Persistent Threat}
  \acro{C2}{command-and-control}
  \acro{CoT}{Chain-of-Thought}
  \acro{CT}{Cyber Twin}
  \acro{CTI}{Cyber Threat Intelligence}
  \acro{DDoS}{Distributed Denial of Service}
  \acro{DT}{Digital Twin}
  \acro{DSR}{defense success rate}
  \acro{DPO}{direct preference optimization}
  \acro{EDR}{Endpoint Detection and Response}
  \acro{GNS3}{Graphical Network Simulator 3}
  \acro{ICS}{Industrial Control System}
  \acro{I2C}[I$^2$C]{Insight Importance Checklist}
  \acro{IDS}{Intrusion Detection System}
  \acro{IoC}{Indicators of Compromise}
  \acro{IM}{Incident Management}
  \acro{IR}{Incident Report}
  \acro{IPS}{Intrusion Prevention System}
  \acro{KG}{Knowledge Graph}
  \acro{LLM}{large language model}
  \acro{ME}{Mitigation Effectiveness}
  \acro{ML}{Machine Learning}
  \acro{NLP}{Natural Language Processing}
  \acro{OCE}{On Call Engineer}
  \acro{PE}{Prompt Engineering}
  \acro{RAG}{Retrieval-augmented generation}
  \acro{SIEM}{Security Information and Event Management}
  \acro{SOAR}{Security Orchestration Automation and Response}
  \acro{SOC}{Security Operations Center}
  \acro{TTP}{Tactics, Techniques, and Procedures}
  \acro{PT}{Penetration Testing}
  \acro{ASP}{Attack Success Parity}
  \acro{GRPO}{Group Relative Policy Optimization}
  \acro{SFT}{Supervised Fine-Tuning}
  \acro{RLVR}{Reinforcement Learning from Verifiable Rewards}
  \acro{POMDP}{Partially Observable Markov Decision Process}
  \acro{PCAP}{packet capture}
  \acro{PPO}{Proximal Policy Optimization}
  \acro{RL}{reinforcement learning}
  \acro{RLAIF}{reinforcement learning from AI feedback}
  \acro{dDPO}{distilled direct preference optimization}
  \acro{dRLAIF}{distilled reinforcement learning from AI feedback}
\end{acronym}